\documentclass[12pt]{article} 
\pdfoutput=1
\usepackage{jheppub}

\usepackage{amsmath,amssymb,amsfonts}

\usepackage{tikz}
\usepackage{dsfont}
\usepackage{comment}

\usepackage{booktabs}
\usepackage{multirow}

\usepackage{xcolor}
\usepackage{graphicx,subcaption}
\usepackage{float}
\usepackage{array,multirow,booktabs,longtable}
\usepackage{bbm}

\DeclareMathOperator{\Tr}{Tr}

\makeatletter\def\l@subsubsection#1#2{}%
\makeatother

\def\be{\begin{eqnarray}}
\def\ee{\end{eqnarray}}

\begin{document}

\setcounter{table}{0}

\title{Charges and holography in 6d (1,0) theories}

\author[a]{Oren Bergman,}
\emailAdd{bergman@physics.technion.ac.il}
\author[a,d,e]{Marco Fazzi,}
\emailAdd{marco.fazzi@mib.infn.it}
\author[b,c]{Diego Rodr\'iguez-G\'omez,}
\emailAdd{d.rodriguez.gomez@uniovi.es}
\author[d,e]{Alessandro Tomasiello}
\emailAdd{alessandro.tomasiello@mib.infn.it}

\affiliation[a]{Department of Physics, Technion, Israel Institute of Technology\\
32000 Haifa, Israel\\[-4mm]}

\affiliation[b]{Department of Physics, Universidad de Oviedo \\ Calle Federico Garc\'ia Lorca 18, E-33007 Oviedo, Spain\\[-4mm]}

\affiliation[c]{Instituto Universitario de Ciencias y Tecnolog\'ias Espaciales de Asturias (ICTEA) \\ Calle de la Independencia 13, E-33004 Oviedo, Spain.\\[-4mm]}

\affiliation[d]{Dipartimento di Fisica, Universit\`a di Milano--Bicocca \\ Piazza della Scienza 3, I-20126 Milano, Italy\\[-4mm]}

\affiliation[e]{INFN, sezione di Milano--Bicocca \\ Piazza della Scienza 3, I-20126 Milano, Italy\\[-4mm]}

\abstract{We study the recently proposed $AdS_7$/CFT$_6$ dualities for a class of 6d ${\cal N}=(1,0)$ theories that 
flow on the tensor branch to long linear quiver gauge theories.
We find a precise agreement 
in the symmetries and in the spectrum of charged states
between the 6d SCFTs and their conjectured $AdS_7$ duals.
We also confirm a recent conjecture that a discrete $S_N$ symmetry relating the baryons in the quiver theories is in fact gauged.}

\maketitle

\section{Introduction and summary}
\label{sec:intro}

The study of quantum field theory becomes harder as one increases the dimension of spacetime: familiar interacting models become non-renormalizable and require new UV completions. However when such completions can be found, they provide interesting windows on non-perturbative physics, whose lessons can be useful in lower dimensions too.
For supersymmetric models, the renormalization group (RG) fixed points at high energies should be superconformal field theories (SCFTs), which can only exist up to $d=6$. Over the years, many realizations for such theories have been proposed using various string-theoretic techniques.

In this paper we will be interested 
in a class of 6d SCFTs with $(1,0)$ supersymmetry \cite{Hanany:1997gh,Brunner:1997gf}, which have effective descriptions in terms of linear quiver gauge theories with gauge group $\Pi_i SU(r_i)$, matter fields charged in bi-fundamental representations, and tensor fields.  Based on their brane engineering, these SCFTs were suggested in \cite{Gaiotto:2014lca} to have $AdS_7$ holographic duals in Type IIA
supergravity, which have been classified \cite{Apruzzi:2013yva} and written down explicitly \cite{Apruzzi:2015wna,Cremonesi:2015bld}. A concrete check of this duality was performed in \cite{Cremonesi:2015bld,Apruzzi:2017nck}, where the $a$ anomaly was compared on both sides and found to agree in full generality. While that test was reassuring, one would like to have further quantitative tests to probe in more detail the non-perturbative physics of these SCFTs.\footnote{Further tests of different aspects were performed in 
\cite{DeLuca:2018zbi,Nunez:2018ags,Filippas:2019puw,Lozano:2019ywa,Lozano:2019jza}.} 

Here we will compare in detail the symmetries and the spectrum of charged operators. It has been known already that the non-abelian symmetries are realized on the 
D8-branes and D6-branes present in the gravity solutions, but the $U(1)$ symmetries are more subtle. On the field theory side, they should acquire masses through a St\"uckelberg mechanism \cite{Hanany:1997gh}; but some of them are in fact anomalous (the analysis becoming particularly interesting in presence of $SU(2)$ gauge groups).
On the gravity side, these $U(1)$'s are particular combinations of the RR gauge field and of the gauge fields that live on branes; we match them with the non-anomalous combinations on the field theory side. 
Amusingly, the matching involves a St\"uckelberg mechanism on the supergravity side.

There are in general two types of charged operators in the field theories we consider.
The first are composed as strings of matter fields which we refer to as {\em string-mesons}, and the second consist of antisymmetrized products of matter
fields which we refer to as {\em baryons}.
These operators carry non-abelian flavor symmetry charges as well as $U(1)$ charges.
Interestingly, the different baryons all have the same charges and dimension. 
The baryons and string-mesons are expected to be related by chiral-ring relations. On the gravity side, we identify the string-mesons with strings connecting the D-branes in the solution, and the baryons with D0-branes. We also provide a realization of the chiral-ring relation through an instantonic process: namely, D0-branes can turn into fundamental strings (and vice versa) via the nucleation of a Euclidean D2-brane which wraps the internal space of the dual gravity background.

When the D0-brane is in a region with non-zero Romans mass, it develops a tadpole that should be canceled by strings ending on it; this corresponds nicely with the structure of the field theory baryons, which are built from rectangular matrices and hence need to include sequences of fields charged under neighboring gauge groups. Taking this into account, however, we find that the D0-brane mass has a single minimum. This suggests that the field theory baryons should all be identified in the SCFT, in stark contrast with most theories in lower dimensions. This is consistent with their charges being equal, and with a recent conjecture \cite{Hanany:2018vph} that these SCFTs have an $S_N$ gauge symmetry. Below we will provide a field-theoretic argument supporting this conjecture which exploits the fact that, for 6d theories with an M5-brane origin, the $S_N$ symmetry can be identified with a subgroup of the gauge group in a certain duality frame (or, in other words, upon performing a different reduction to Type IIA). Furthermore, we are able to match exactly and in general the charges and dimension of this single baryon to the mass of the D0-brane (with strings attached) at its minimum.

This paper is organized as follows. In section \ref{sec:field} we present the gauge theory description of the 
6d $(1,0)$ SCFTs, highlighting their global symmetries and spectrum of gauge-invariant charged BPS operators. 
In section \ref{sec:sugra} we review the  construction of the $AdS_7$ gravity duals; we identify the non-abelian gauge fields as well as the non-anomalous abelian ones coming from the reduction of the ten-dimensional Type IIA theory on the internal space; we construct the charged states and compute their masses and charges, matching them with those of the dual operators. 
In sections \ref{sec:field} and \ref{sec:sugra} we concentrate on the case in which the 6d quiver gauge theory
has a segment in which the gauge group ranks do not change, namely a plateau, and the dual Type IIA background has a region with
a vanishing Romans mass.
In section \ref{PlateauLess} we briefly discuss the case in which the gauge theory does not have a plateau, and the dual Type IIA background
does not have a massless region, mainly emphasizing the differences with the case containing a plateau.
In section \ref{Examples} we work out three explicit examples, including a plateau-less case. 
In two appendices we include a supersymmetry analysis of D0-branes and strings, and some details related to 
the computation of their $U(1)$ charges.

\section{Field theory}
\label{sec:field}

\subsection{Linear quivers}

\begin{figure}[th!]
\centering
\includegraphics[scale=.6]{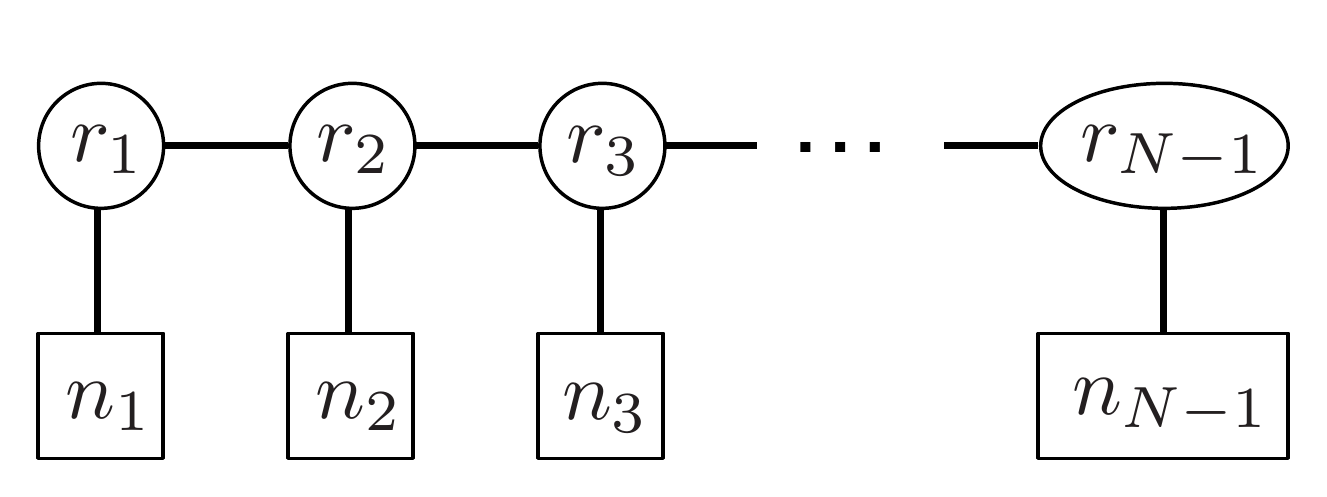} 
\caption{The general 6d linear quiver theory.} 
\label{QuiverGeneral}
\end{figure}

\begin{figure}[ht!]
\centering
\includegraphics[scale=.35]{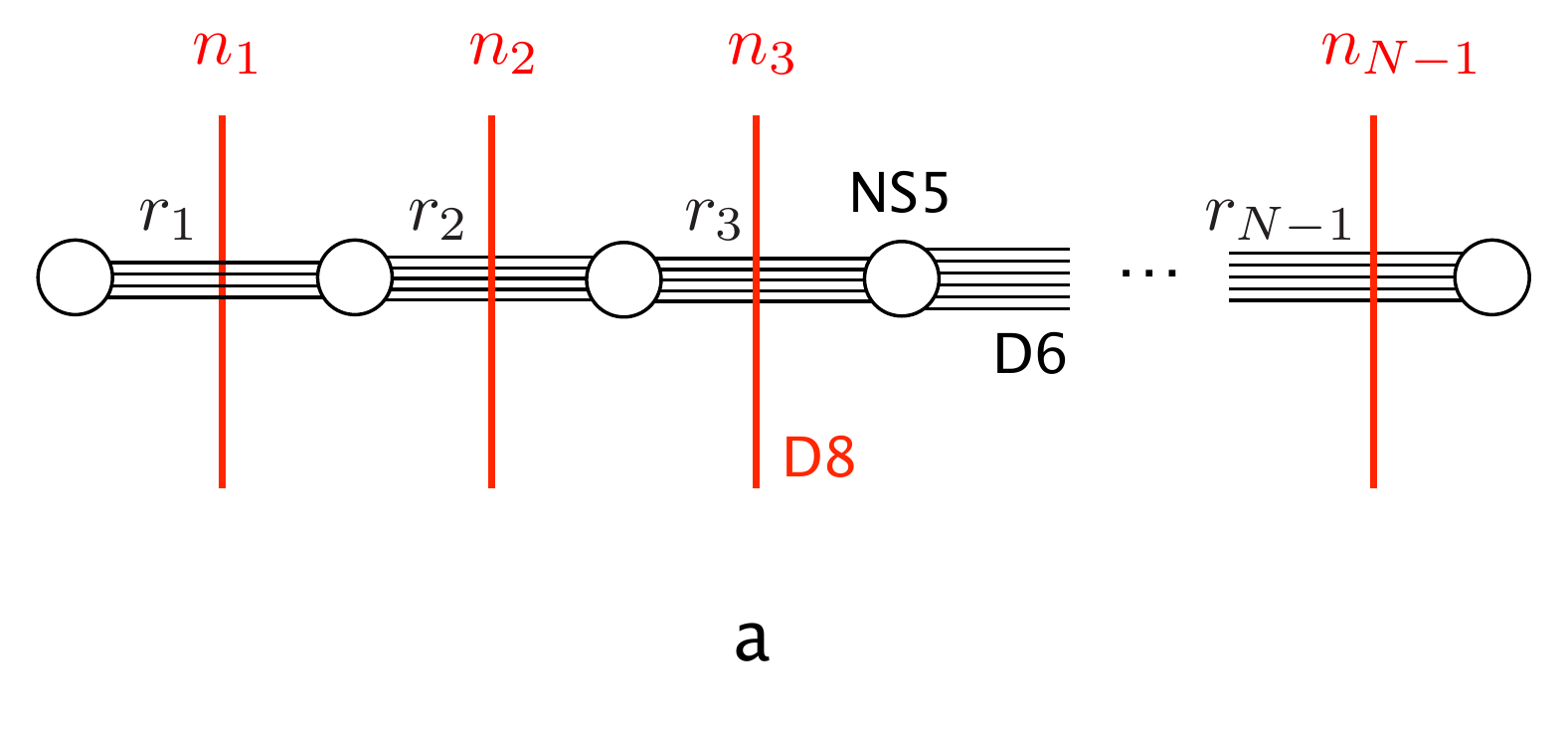} 
\hspace{20pt}
\includegraphics[scale=.35]{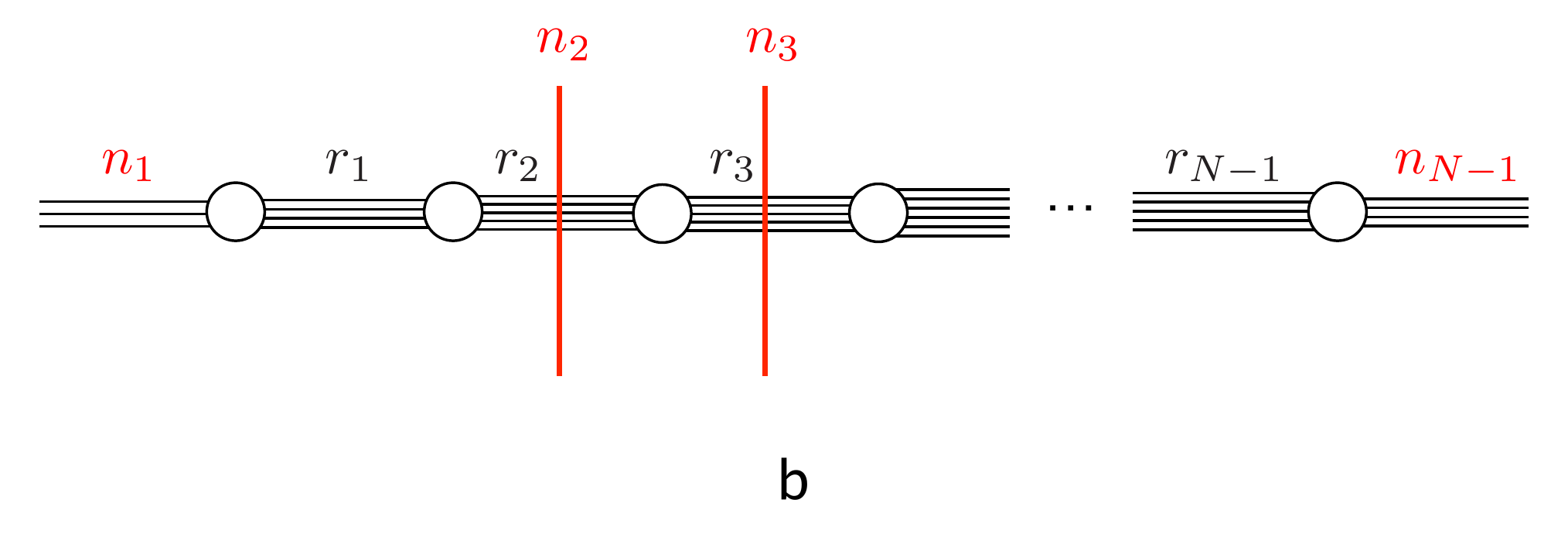} 
\caption{Brane configurations for linear quivers. In (b) the first and last stacks of flavor D8-branes of (a) are replaced by semi-infinite D6-branes.} 
\label{BranesGeneral}
\end{figure}

The general structure of the theories we are interested in is shown in Fig.~\ref{QuiverGeneral}.
These are linear quiver gauge theories with $N-1$ gauge nodes $SU(r_i)$, where $i=1,\ldots, N-1$, a bi-fundamental hypermultiplet for
each $SU(r_i)\times SU(r_{i+1})$ pair, and $n_i$ fundamental flavor hypermultiplets for $SU(r_i)$.
These theories are free of gauge anomalies provided that each gauge group $SU(r_i)$ has effectively $2r_i$ flavors, namely that
\be
\label{AnomalyCondition}
n_i = 2r_{i} - r_{i +1} - r_{i -1} \,.
\ee
There exists a simple brane construction for this theory in Type IIA string theory
\cite{Brunner:1997gf, Hanany:1997gh}.
It consists of $N$ parallel NS5-branes separated along one coordinate, 
with $r_i$ D6-branes stretched between the $i$th NS5-brane and the $(i+1)$st NS5-brane,
and a stack of $n_i$ D8-branes intersecting the D6-branes between the $i$th and $(i+1)$st NS5-brane, Fig.~\ref{BranesGeneral}a.
The first and last D8-brane stacks can be traded for semi-infinite D6-branes via a Hanany-Witten transition, as shown in Fig.~\ref{BranesGeneral}b.
For each gauge node there is also a tensor multiplet corresponding to the NS5-brane degrees of freedom.
The anomaly cancellation condition (\ref{AnomalyCondition}) is seen as a tadpole cancellation condition on the NS5-brane worldvolumes.

\begin{figure}[ht!]
\centering
\includegraphics[scale=.6]{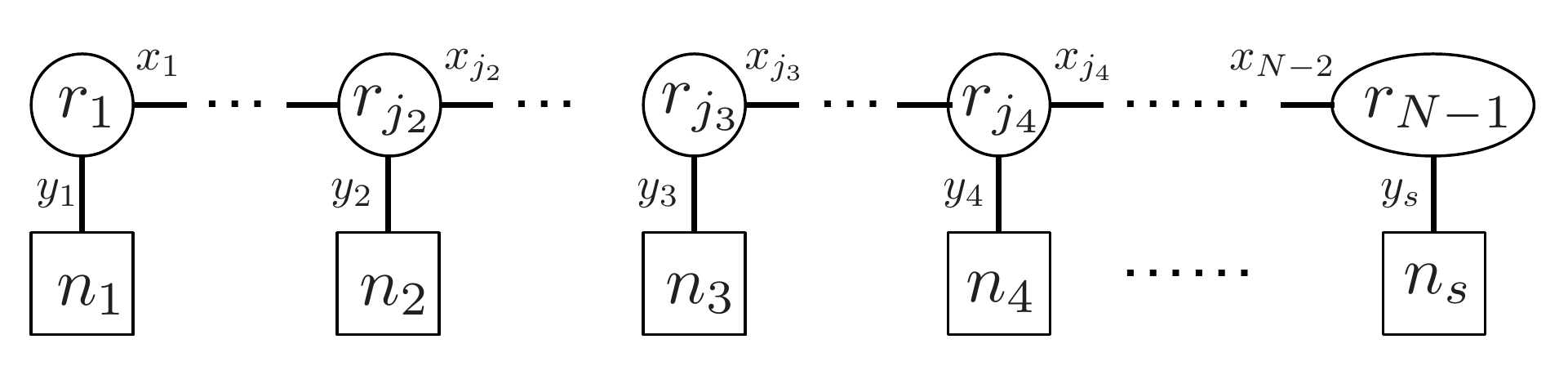} 
\caption{The general 6d linear quiver theory with $s$ sets of flavors. We denote by $x_i,\tilde{x}_i$ the two complex scalar components of the 
bi-fundamental hypermultiplets,
and by $y_a,\tilde{y}_a$ the two complex scalar components of the fundamental hypermultiplets.} 
\label{QuiverGeneral2}
\end{figure}

For our purposes we will assume that the flavored gauge nodes are well separated, so
that the quiver theories take the form of Fig.~\ref{QuiverGeneral2}.
We label the flavors by $b=1,\ldots,s$, and the flavored nodes are labelled by $j_b$.
We assume that the first and last nodes are flavored, namely that $j_1=1$ and $j_s=N-1$.
The anomaly condition (\ref{AnomalyCondition}) implies that $r_j$ is a convex function of $j$, and therefore that
it either has a maximal plateau in some segment between consecutive flavors $j_p\leq j \leq j_{p+1}$,
or that it is maximized at one of the ends of the quiver.
In the remainder of this section, as well as in the next section, we will assume that there exists a plateau between the $p$th and $(p+1)$st flavors.
In section \ref{PlateauLess} we will briefly summarize the results for the plateau-less case.

The anomaly condition (\ref{AnomalyCondition}) can be solved by imposing the ``boundary conditions" $r_0=r_N=0$, to give
\be
\label{RanksPlateau}
r_j = \left\{
\begin{array}{ll}
r_\text{max} - \sum_{b=a_\text{R}(j)}^{p} (j_b - j) n_b  & \; \text{for} \; j \leq j_p \\
r_\text{max} & \; \text{for} \; j_{p} \leq j \leq j_{p+1} \\
r_\text{max} - \sum_{b=p+1}^{a_\text{L}(j)} (j - j_b) n_b  & \; \text{for} \; j \geq j_{p+1} \,,
\end{array}
\right.
\ee
where $a_{\text{R},\text{L}}(j)$ labels the flavor nearest the $j$th node on the right and left, respectively, and $r_\text{max}$ is the maximal gauge group rank, which is given by
\be
\label{MaxRankPlateau}
r_\text{max} = \sum_{b=1}^p j_b n_b = \sum_{b=p+1}^s (N-j_b) n_b \,.
\ee

\subsection{Global symmetry} 
\label{GlobalSymmetry}

The gauge nodes are $SU(r_i)$ rather than $U(r_i)$, since the
the $U(1)$ gauge bosons acquire a mass via a St\"uckelberg coupling to a scalar in the tensor multiplet \cite{Hanany:1997gh}.
The $U(1)$ symmetries are therefore global symmetries in the low energy theory.
However in general they are anomalous.
Let us denote by $J^\mu_i$, with $i=1,\ldots, N-2$, the $U(1)$ currents associated to the bi-fundamental fields $x_i$,
and by $I^\mu_b$, with $b=1,\ldots,s$, the $U(1)$ currents associated to the flavor fields $y_b$.
The anomalies are given by
\be
\label{Anomaly1}
\partial_\mu J_i^\mu &=& r_i \Tr(F_{i+1}^3) -  r_{i+1} \Tr(F_i^3) \\
\partial_\mu I_b^\mu &=& n_b \Tr(F_{j_b}^3) \,.
\ee
Anomaly-free currents are found by summing from flavor to flavor with appropriate coefficients.
There are $s-1$ independent conserved currents given by summing from the $b$th flavor
to the $(b+1)$st flavor as follows (up to an overall normalization)
\be
\label{AnomalyFreeSymmetries}
J_{b}^\mu = \frac{n_{b+1}}{r_{j_b}} I_b^\mu + \sum_{j=j_b}^{j_{b+1}-1} \frac{n_b n_{b+1}}{r_j r_{j+1}} J_j^\mu
- \frac{n_b}{r_{j_{b+1}}} I_{b+1}^\mu\,.
\ee
The global symmetry of the 6d theory is therefore generically $U(1)^{s-1} \times \prod_{b=1}^s SU(n_b)$.\footnote{For an F-theory perspective on the global symmetry of these SCFTs, with particular emphasis on the $U(1)$ factors, see \cite{Apruzzi:2020eqi}.
}
In the next section we will see that this agrees with the gauge symmetry of the proposed dual $AdS_7$ background.

\subsubsection{$SU(2)$ node}

\begin{figure}[ht!]
\centering
\includegraphics[scale=.6]{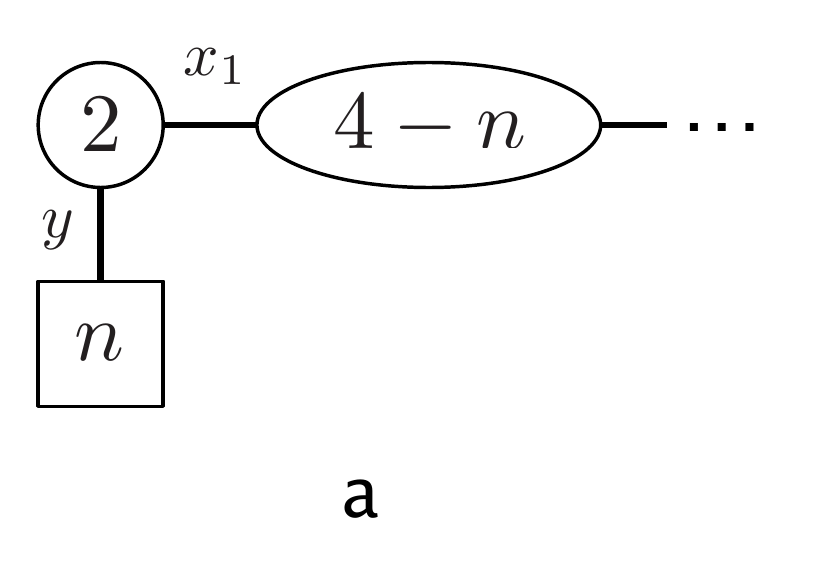}
\hspace{20pt}
 \includegraphics[scale=.6]{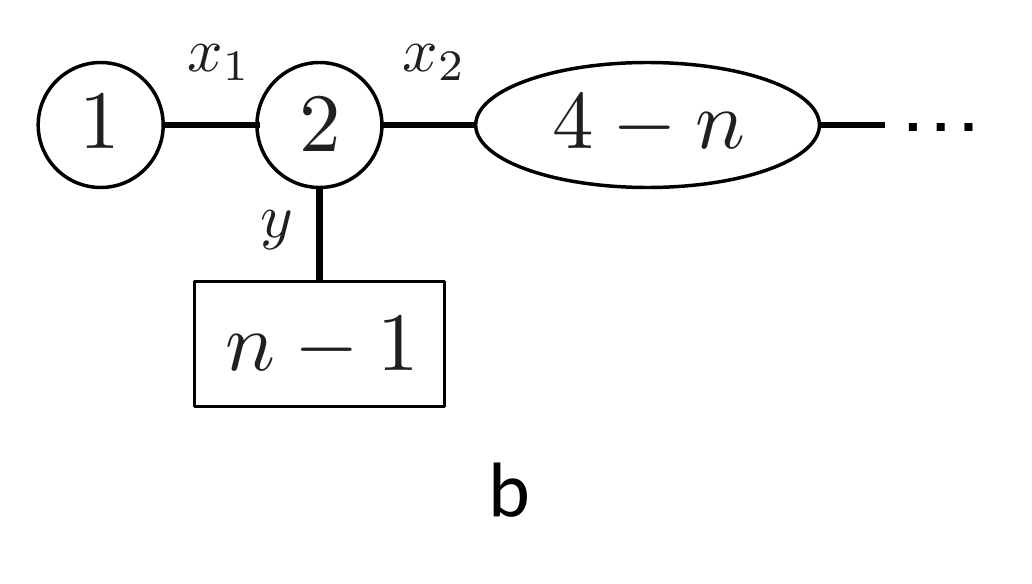}
\caption{An $SU(2)$ piece of the quiver.}
\label{QuiverSU(2)}
\end{figure}

\begin{figure}[ht!]
\centering
\includegraphics[scale=.6]{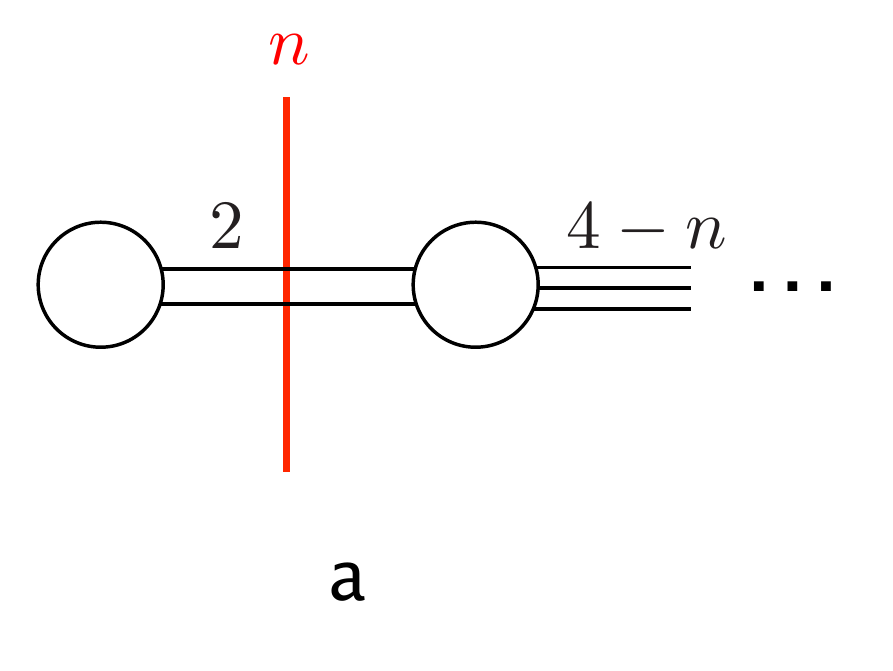} 
\hspace{20pt}
\includegraphics[scale=.6]{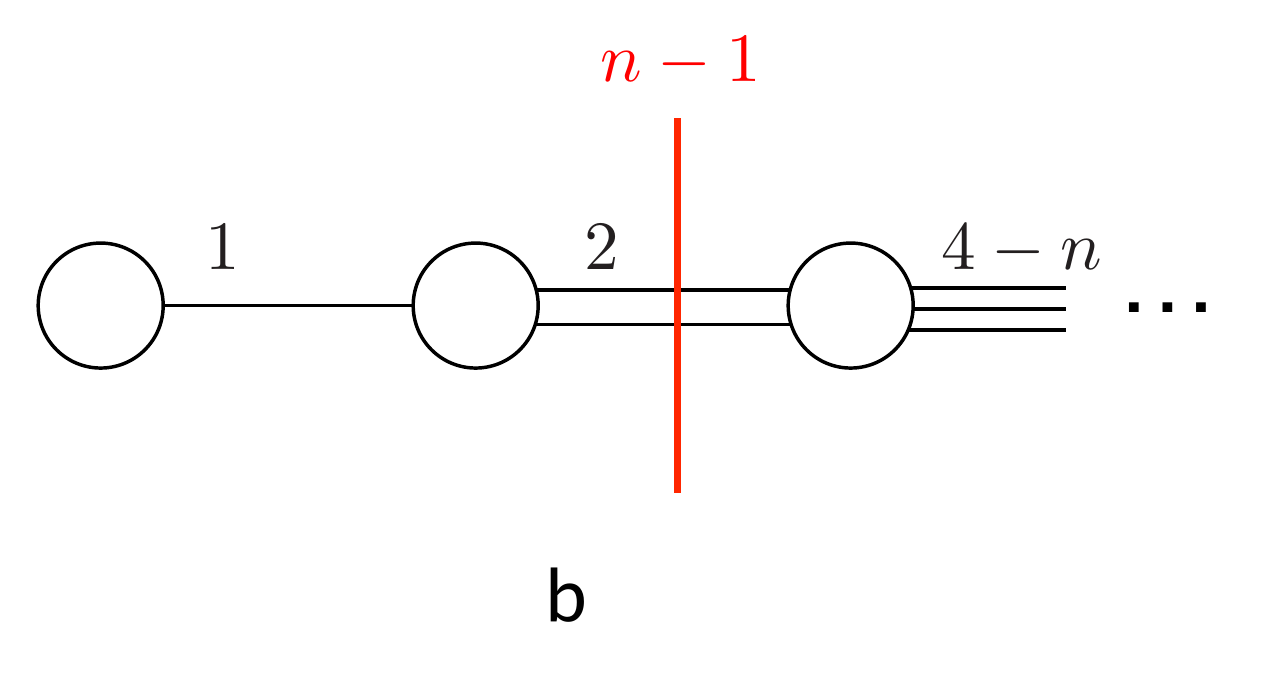} 
\caption{Brane configurations for the quiver tails in Fig.~\ref{QuiverSU(2)}.} 
\label{BranesSU(2)}
\end{figure}

There are a number of special cases that deserve our attention.
These occur when there is an $SU(2)$ factor.
The $U(1)$ anomaly is absent for $SU(2)$ since $\text{Tr}F^3_{SU(2)} = 0$.
This would suggest that there are additional $U(1)$ symmetries in these cases.
However this is not the case as we will now explain.
The most general situations with an $SU(2)$ factor are depicted in Fig.~\ref{QuiverSU(2)},
which shows an edge of the quiver containing an $SU(2)$ gauge node with $n$ flavors, where $n=0,\ldots, 4$.
There are actually two different theories with this gauge field and matter content. 
Fig.~\ref{QuiverSU(2)}a shows the minimal theory with $n$ flavors, and Fig.~\ref{QuiverSU(2)}b shows the theory
which has an extra tensor multiplet associated to the edge.
This is represented as an ``$SU(1)$" gauge node which is really a flavor.
The corresponding brane configuration is shown in Fig.~\ref{BranesSU(2)}b, where one of the D8-branes is traded for an extra
NS5-brane on which a single D6-brane ends.
In the first case the $U(1)$ symmetry associated to $y$ is anomaly-free by itself, and in the second case
the two $U(1)$ symmetries associated to $x_1$ and $y$ are anomaly-free by themselves.
However these symmetries are actually absent in the SCFT.

Let us begin with the edge without the extra tensor multiplet, Fig.~\ref{QuiverSU(2)}a.
The relevant cases are $n=0, 1 , 2$ and $4$. 
The case with $n=3$ is equivalent to the case with $n=4$ of Fig.~\ref{QuiverSU(2)}b, which we will discuss below.
For $n=4$ this is just an $SU(2)$ theory with 4 flavors.
The classical global symmetry of this theory is $SO(8)$.
However it has been argued that the symmetry is reduced in the SCFT to $SO(7)$ \cite{Ohmori:2015pia,Hanany:2018vph}.
The other cases can be obtained by gauging an $SU(4-n)$ subgroup of the global symmetry.
Classically this would leave a $U(1)$ global symmetry acting on 
the bi-fundamental field $x_1$ for $n=0$, and two $U(1)$ global symmetries acting on $x_1$ and on the flavor $y$ for $n=1$.
For $n=2$ it would leave an $SU(2)^3$ global symmetry, with two $SU(2)$ factors acting on the flavors and one acting on the bi-fundamental field.
However since the global symmetry of the $n=4$ theory is actually $SO(7)$ these conclusions are modified.
For $n=0$ the edge has no global symmetry, for $n=1$ it only has a single $U(1)$ corresponding precisely 
to the first two terms in (\ref{AnomalyFreeSymmetries}), and for $n=2$ it has $SU(2)^2$.\footnote{In more detail
$SO(8) \supset SU(4)\times U(1)_1 \supset SU(3)\times U(1)_1\times U(1)_2$ with ${\bf 8}_v \rightarrow {\bf 3}_{0,3} + {\bar{\bf 3}}_{0,-3}
+ {\bf 1}_{1,0} + {\bf 1}_{-1,0}$, whereas $SO(7)\supset SU(4) \supset SU(3)\times U(1)$ with
${\bf 8}_s \rightarrow {\bf 3}_{3} + {\bar{\bf 3}}_{-3}
+ {\bf 1}_{3} + {\bf 1}_{-3}$.}
We conclude that the formula for the anomaly-free $U(1)$ currents (\ref{AnomalyFreeSymmetries}) holds also in the cases
with an $SU(2)$ edge of the form shown in Fig.~\ref{QuiverSU(2)}a with $n=0,1$, and that there are no extra $U(1)$ symmetries.
For $n=2$ the full quiver is actually fixed to be $2+SU(2)^{N-1}+2$.
The classical global symmetry is $SU(2)^{N+2}$, but by repeating the above analysis at each of the nodes one can show that this reduces to just 
$SU(2)^3$ \cite{Hanany:2018vph}.

Now consider the $SU(2)$ edge with the extra tensor multiplet, Fig.~\ref{QuiverSU(2)}b.
Let us start with the case $n=4$, namely an $SU(2)$ theory with 4 flavors and an extra tensor multiplet.
It has been argued that in this case the global symmetry is reduced to $G_2$ \cite{Hanany:2018vph}.
As before the other cases correspond to gauging an $SU(4-n)$ subgroup, now with $n=1,2$.\footnote{For $n=3$ this is an $SU(2)$ theory with 4 flavors
and two extra tensor mutiplets. This theory has a global symmetry $SU(3)$.}
For $n=1$ we get no remaining global symmetry, and for $n=2$ we are left with $SU(2)$.\footnote{In more detail, $G_2\supset  SU(3)$ with
${\bf 7} \rightarrow {\bf 3} + {\bar{\bf 3}} +{\bf 1}$, and $G_2\supset  SU(2)\times SU(2)$ with
${\bf 7} \rightarrow ({\bf 2},{\bf 2}) + ({\bf 1},{\bf 3})$.}
So, again, there are no extra $U(1)$ symmetries associated to the $SU(2)$ edge.
The only $U(1)$ symmetries are those given in (\ref{AnomalyFreeSymmetries}).

\subsection{Charged operators}

Our main interest is in the spectrum of gauge-invariant charged BPS operators.
There are two types of such operators in the theory.
The first are given by strings of matter fields beginning and ending with a flavor field, which we will refer to as {\em string-mesons}.
There are $s-1$ independent string-mesons which we can take as beginning at the $b$th flavor and ending at the $(b+1)$st flavor:
\be
\label{StringMesonGeneral}
{\cal M}_{b} = y_b \cdot \left( \prod_{j=j_b}^{j_{b+1}-1} x_j\right) \cdot \tilde{y}_{b+1} \,.
\ee
This operator has a  scaling dimension $\Delta_{\mathcal{M}_b} = 2(j_{b+1}-j_b+2)$, and transforms in the 
bi-fundamental representation of $SU(n_b)\times SU(n_{b+1})$. From (\ref{AnomalyFreeSymmetries})
we also deduce that it carries charges under at most three of the $U(1)$ symmetries:
\be
\label{StringMesonCharges}
Q_c(\mathcal{M}_b) =\begin{cases} \displaystyle
 - \frac{n_{b-1}}{r_{j_b}}  & \text{for}\ c=b-1 \\
\displaystyle  \frac{n_{b+1}}{r_{j_b}} + \sum_{k=j_b}^{j_{b+1}-1} \frac{n_b n_{b+1}}{r_k r_{k+1}} + \frac{n_b}{r_{j_{b+1}}}  &\text{for}\ c=b \\
 \displaystyle - \frac{n_{b+2}}{r_{j_{b+1}}} \displaystyle & \text{for}\ c=b+1\\
 0 & \text{otherwise}
 \end{cases}
 \ee
 The first charge is zero for $b=1$, whereas the third is zero for $b=s-1$.
 
The second type of operators are given by antisymmetrized products.
We will therefore refer to them as {\em baryons}.
There are $N$ independent baryon operators ${\cal B}_j$ with $j=0,\ldots, N-1$, given by
\be
\label{BaryonPlateau}
{\cal B}_j  = \begin{cases} 
\displaystyle(x_j)^{r_j} \cdot \prod_{b=a_\text{R}(j)}^{p}\left[\left(\prod_{i=j+1}^{j_b-1} \tilde{x}_i\right)\cdot y_b\right]^{n_b} & \text{for} \; 0\leq j \leq j_{p}-1 \\
(x_j)^{r_\text{max}} = \det x_j & \text{for} \; j_{p} \leq j \leq j_{p+1} - 1 \\
\displaystyle (x_j)^{r_{j+1}} \cdot \prod_{b=p+1}^{a_\text{L}(j+1)}\left[ \tilde{y}_b \cdot \left(\prod_{i=j_b}^{j-1} \tilde{x}_i\right)\right]^{n_b} & \text{for} \; j_{p+1} \leq j \leq N-1
\end{cases}
\ee
where all products are antisymmetrized. 
All of these have the same scaling dimension $\Delta_{{\cal B}_j} = 2r_\text{max}$, they are all singlets under the non-abelian part 
of the global symmetry, and all carry only $U(1)_p$ charge:
\be
\label{BaryonChargesPlateau}
Q_c({\cal B}_j) = \left\{
\begin{array}{ll}
 \displaystyle \frac{n_{p} n_{p+1}}{r_\text{max}} & \; \text{for} \; c=p \\
0 & \; \text{for} \; c\neq p \,.
\end{array}
\right.
\ee

The string-meson operators and the baryon operators are not completely independent.
The product of the $N$ baryons can be expressed in terms of the $s-1$ string-mesons in a chiral-ring-like relation:
\be
\label{ChiralRingPlateau}
\prod_{j=0}^{N-1} {\cal B}_j \sim 
\left[\prod_{b=1}^{p-1} ({\cal M}_b \cdots {\cal M}_{p-1})^{n_b j_b}\right] \cdot ({\cal M}_p)^{r_\text{max}} \cdot
\left[\prod_{b=p+1}^{s-1} ({\cal M}_{p+1} \cdots {\cal M}_b)^{n_{b+1} (N-j_{b+1})}\right] \,, \nonumber \\
\ee
with appropriate antisymmetrization of the products on the RHS.
Here we are omitting neutral mesonic factors of the form $x_i \tilde{x}_i$ and $y_b \tilde{y}_b$,
and presenting only the charged components of the chiral-ring relation.
Note in particular that the RHS is a singlet under all the non-abelian factors of the global symmetry.
The first term on the RHS transforms as 
$\left(\bar{\bf n}_{p}\right)^{r_\text{max} - n_p j_p}_\text{asym}$ of $SU(n_{p})$,
the second term as $\left({\bf n}_p,\bar{\bf n}_{p+1}\right)^{r_\text{max}}_\text{asym}$ of $SU(n_p)\times SU(n_{p+1})$,
and the third term as
$\left({\bf n}_{p+1}\right)^{r_\text{max} - n_{p+1}(N-j_{p+1})}_\text{asym}$ of $SU(n_{p+1})$.
Antisymmetrizing the product of these gives a singlet.

\subsection{A discrete (gauge) symmetry}

In addition to the continuous global symmetries discussed in section~\ref{GlobalSymmetry}, the linear quiver theories 
also appear to have an emergent discrete symmetry $S_N$, originally identified in 
\cite{Hanany:2018vph}.\footnote{The authors of \cite{Hanany:2018vph} considered a special class of quiver theories corresponding
to M5-brane theories, but their arguments are easily extended to the more general quiver theories considered here.}
This symmetry is not seen at the level of the action, but
becomes apparent at the level of the gauge-invariant operators discussed above.
Since the $N$ baryon operators ${\cal B}_i$ all have the same scaling dimensions and charges under the continuous global symmetries,
there is an $S_N$ symmetry that permutes them.
From the point of view of the Type IIA brane construction this corresponds to the permutation of the $N$ NS5-branes.

It has been conjectured in \cite{Hanany:2018vph} that this symmetry is actually gauged.
This was argued using an auxiliary three-dimensional $\mathcal{N}=4$ quiver gauge theory, the so-called {\em magnetic quiver},
whose Coulomb branch is conjectured to be identical to the Higgs branch of the 6d {\em electric quiver}.
The Coulomb branch of the three-dimensional theory is given by the closure of a certain nilpotent orbit of 
the global symmetry. 
When the $N$ NS5-branes (or M5-branes) are brought together, the new nilpotent orbit is the $S_N$ quotient of the one for the separated branes.

There is another way to see that the $S_N$ symmetry is gauged, at least for the 6d theories on M5-branes, as we will now explain.
Some, but not all, 6d SCFTs are realized in terms of M5-branes in various M-theory backgrounds.
These theories have two different gauge theory deformations.
The first is a 6d quiver gauge theory of the type we have been discussing, corresponding to the low energy effective 
theory on the tensor branch of the 6d SCFT.
From the point of view of string/M-theory this corresponds to separating the M5-branes and reducing along a direction transverse to them.
For example, for the 6d $(2,0)$ theory of $N$ M5-branes in flat space this gives a Type IIA brane configuration with a single infinite D6-brane
intersecting $N$ NS5-branes, namely a theory of $N$ free hypermultiplets and $N$ free tensor multiplets.
As we explained above, this description exhibits an emergent global $S_N$ symmetry.
The second gauge theory deformation is a 5d supersymmetric gauge theory obtained by compactifying the 6d SCFT on a circle,
namely reducing along one of the M5-brane directions.
This is the theory living on $N$ D4-branes in Type IIA string theory in a background corresponding to the reduction of the M-theory background.
For example the reduction of the $(2,0)$ theory gives the 5d ${\cal N}=2$ SYM theory with gauge group $U(N)$.
In this case the permutation group $S_N$ is part of the gauge symmetry: it is the Weyl group of $U(N)$.
While the two gauge theories are clearly different, they both correspond to deformations of the same 6d SCFT.
We conclude from this that the $S_N$ symmetry of the 6d quiver theory is in fact gauged, at least in the case of M5-brane theories.
While this argument is not directly applicable to 6d quiver theories that are not deformations of M5-brane theories,
since these do not have an obvious 5d gauge theory reduction, it is plausible that the conclusion continues to hold, 
given that these theories can often be related to ones that are deformations of M5-brane theories via Higgsing
(see for example section \ref{sub:symflav}).

We will confirm that $S_N$ is gauged in all of these theories using holography.
A gauged $S_N$ implies that there is only one physical baryon operator corresponding to the sum of the $N$ baryons.
Our analysis will show that there is indeed a unique bulk state dual to this operator.

\section{Supergravity}
\label{sec:sugra}

\subsection{Solutions}
\label{sub:review}

We begin this section by briefly reviewing the
Type IIA $AdS_7$ solutions of \cite{Apruzzi:2013yva} that are conjectured to be dual to 6d $(1,0)$ theories of the type 
described in the previous section 
\cite{Gaiotto:2014lca,Apruzzi:2015zna,Apruzzi:2015wna}. 
The ten-dimensional geometry is a warped product $AdS_7\times M_3$, with the internal space $M_3$ 
having the topology of $S^3$, given by an $S^2$ fibered over an interval.
There is NSNS 3-form flux on $M_3$ and RR 2-form flux on the $S^2$ fiber, and in general also
a non-vanishing RR 0-form flux $F_0$, the so-called Romans mass.
For $F_0=0$ the Type IIA background lifts
to the M-theory background $AdS_7\times S^4/\mathbb{Z}_k$.

The Type IIA  background is defined by a single piecewise smooth function on the interval $\alpha(z)$,
with $z\in[0,N]$, satisfying the differential equation\footnote{The first-order supergravity equations reduce to this single ODE.}
\be 
\label{eq:romans}
\dddot{\alpha}(z) = - 2\pi(9\pi)^2 F_0(z) \,,
\ee
with appropriate boundary conditions corresponding to the asymptotics of the brane configuration.
Let us divide the interval into $N$ equal segments with $j\leq z \leq j+1$, where $j=0,\ldots, N-1$, corresponding to the $N$ NS5-branes in the Type IIA brane construction.
The Romans mass $F_0(z)$ is a piecewise constant function given in the segment $z \in (j,j+1)$ by
\be
\label{RomansMass}
2\pi F_{0}(z) = r_{j+1} - r_j \,, 
\ee
where $r_j$ is the rank (plus 1) of the $j$th gauge group in the dual quiver theory.
Recall that we impose the boundary conditions $r_0=r_N\equiv 0$.
The D8-branes present in the original brane construction remain as sources localized at $z=j_b$, with $b=1,\ldots,s$,
across which the Romans mass jumps by $2\pi \Delta F_0 = n_b$.
The plateau is the region $j_p \leq z \leq j_{p+1}$ in which $F_0=0$.
We will refer to this as the {\em massless region}.
The case without a massless region will be discussed in the next section.
It follows from (\ref{RomansMass}) and (\ref{RanksPlateau}) that 
\be
\label{RomansMass2}
2\pi F_0(z) = \begin{cases}
- \sum_{b=a_\text{R}(j)}^p n_b & \; \mbox{for} \; j<j_p \\
0 & \; \mbox{for} \; j_p\leq j \leq j_{p+1} \\
\sum_{b=p+1}^{a_\text{L}(j)} n_b & \; \mbox{for} \; j>j_{p+1} \,.
\end{cases}
\ee

The solution to (\ref{eq:romans}) for $j \leq z \leq j+1$ takes the form \cite{Cremonesi:2015bld}
\be
\label{Solution}
\alpha(z) = \alpha_j -2(9\pi) y_j (z-j) -\frac{(9\pi)^2}{2} r_j (z-j)^2 -\frac{(9\pi)^2}{6} (r_{j+1}-r_j) (z-j)^3\ , \label{eq:alphaz}
\ee
where $\alpha_j$ and $y_j$ are fixed in terms of $r_j$ by the boundary and continuity conditions
(see \cite{Cremonesi:2015bld,Apruzzi:2017nck} for the general expressions).
The ten-dimensional metric is explicitly given by
\begin{align}\label{eq:10dmetricz}
	ds^2_{10} &= e^{2A(z)} \,ds^2_{{AdS}_7}+ (4\pi)^2 e^{-2A(z)} \left(dz^2 + \frac{\alpha^2}{\dot \alpha^2 - 2 \alpha \ddot \alpha} \,ds^2_{S^2}\right)\ ,
\end{align}
where $ds^2_{{AdS}_7}$ is the unit-radius metric on $AdS_7$, $ds^2_{S^2}$ is the unit-radius metric on $S^2$ with coordinates $\{\theta,\phi\}$ (and volume form $d\Omega_2 = \sin\theta d\theta \wedge d\phi$), and the warp factor is
\begin{equation}\label{eq:warpingz}
e^{2A(z)} = 2^{7/2}\pi \left(-\frac \alpha{\ddot \alpha}\right)^{1/2}\,.
\end{equation}
The dilaton is given by
\begin{equation}\label{eq:dilatonz}
e^{\phi(z)} 
=\frac{1}{\sqrt{\pi}}\left(\frac{3}{2}\right)^4  \,e^{3A} \,(\dot \alpha^2-2 \alpha \ddot \alpha)^{-1/2}\,.
\end{equation}
The background is in general singular at the two poles of $M_3$, reflecting the
presence of the semi-infinite D6-branes, or equivalently the D8-branes at $z=j_1=1$ and $z=j_s=N-1$.
The gauge-invariant NSNS and RR field strengths are given by
\be
\label{GeneralFieldStrengths}
H_3 &=& \dot{h}(z)\, dz\wedge d\Omega_2 \,,\\ 
{F}_2 & = & dC_1 - F_0 B_2 = f(z) \,d\Omega_2 
\ee
where 
\be
f(z) &=&  \frac{1}{2(9\pi)^2}\left(\ddot{\alpha}- \frac{\alpha \dot{\alpha}\dddot{\alpha}}{\dot \alpha^2-2 \alpha \ddot \alpha}\right)\, ,  \\
h(z) &=& \pi \left( z - \frac{\alpha \dot \alpha}{\dot \alpha^2-2 \alpha \ddot \alpha}\right) \,,
\ee
and the corresponding fluxes are 
\be
\label{NSNSFlux}
\int_{M_3} H_3 &=& 4\pi^2 N \, ,\\
\int_{S^2} {F}_2 &=& 4\pi f(z) \,.
\ee
The latter corresponds to the so-called Maxwell D6-brane charge, which is gauge-invariant but not quantized.

A quantity that is more closely related to the number of D6-branes is the Page charge, defined in general by $\tilde{F} = F\wedge e^{B_2}$,
and given in this case by $\tilde{F}_2 = F_2 + F_0 B_2 = dC_1$. 
Page charge is quantized but not invariant under gauge transformations of the NSNS potential $B_2$.
It is convenient to work in a gauge in which $B_2$ only has components along the $S^2$,
and vanishes at the two poles of $M_3$.
This requires performing a large gauge transformation between the poles of $M_3$ such that 
$B_2 \rightarrow B_2 - \pi N d\Omega_2$.
We will assume that this gauge transformation is performed at a point in the massless region, namely at $z=z_0$ where 
$j_p < z_0 <j_{p+1}$. The gauge 1-form function is therefore given by
$\Lambda_1(z) = \pi N \Theta(z-z_0) \cos\theta d\phi$, and we have
\be
\label{NSNSGaugePlateau}
B_2(z) = 
\begin{cases}
h(z)\, d\Omega_2 & \text{for}\;  z < j_p \\
\pi N \delta(z-z_0) dz \wedge \cos\theta d\phi +  (-\pi N \Theta(z-z_0)+ h(z))\,d\Omega_2 & \text{for}\; j_p<z < j_{p+1} \\
  (-\pi N + h(z))\, d\Omega_2 & \text{for}\; z > j_{p+1} \,. \\
\end{cases} \nonumber \\
\ee
The D6-brane Page charge in the segment $j < z < j+1$  is subsequently given by
\be
\label{RRFluxPlateau}
\int_{S^2} \frac{\tilde{F}_2}{2\pi}  &=&
\begin{cases}
  jr_{j+1} - (j+1)r_j &   \text{for}\;  j \leq j_p -1 \\
 - r_\text{max} &   \text{for}\;  j_p \leq j \leq j_{p+1} -1  \\
  (j-N)r_{j+1} - (j-N+1)r_{j} &  \text{for}\; j \geq j_{p+1} 
\end{cases} \nonumber \\
&=&
\begin{cases}
 - \sum_{b=1}^{a_\text{R}(j)-1} j_b n_b  &   \text{for}\; j \leq j_p -1 \\
 - r_\text{max} &   \text{for}\;   j_p \leq j \leq j_{p+1} -1 \\
 - \sum_{b=a_\text{L}(j)+1}^s (N-j_b) n_b   &  \text{for}\; j \geq j_{p+1} 
\end{cases} 
\ee
where in the second equality we used (\ref{RanksPlateau}) and (\ref{MaxRankPlateau}).
This has a discontinuity across the location of each D8-brane stack at $z=j_b$ that is accounted for by a D6-brane charge carried
by the D8-brane stack itself. This is sourced by a worldvolume magnetic flux, which for a D8-brane in the $b$th stack
is\footnote{This quantity is also not gauge-invariant
under the gauge transformation of $B_2$. 
The gauge-invariant 
combination is $2\pi \mathcal{F}= 2\pi f + B_2|_\text{D8}$. In the gauge (\ref{NSNSGaugePlateau}) this gives (\ref{D8MagneticFlux}).}
\be
\label{D8MagneticFlux}
\int_{S^2} \frac{f_b}{2\pi} = \mu_b = \begin{cases}
  j_b &   \text{for}\;  b \leq p \\
 j_b - N &   \text{for}\;  b \geq p+1 \,.
  \end{cases}
\ee
In particular the D6-brane charges carried by the first D8-brane stack at $z=j_1=1$ and by the last D8-brane stack at $z=j_s=N-1$
account for the singularities at the two poles.
For large $N$ these are effectively at $z=0$ and $z=N$.

\subsection{Symmetries}

As usual the global symmetries of the SCFT should correspond to gauge symmetries in the bulk,
namely to massless vector fields.
The 6d conformal symmetry is of course dual to the isometry of $AdS_7$.
The $SU(2)_R$ symmetry is realized in the bulk as the isometry group of the $S^2$ fiber of $M_3$.
In particular the $U(1)_R$ Cartan subgroup corresponds to the $U(1)$ isometry associated to the 1-form $\cos\theta \, d\phi$.
These are the only symmetries arising from the geometry.
Additional symmetries arise from ten-dimensional supergravity gauge fields and from the source D8-branes.
There is a $U(1)$ gauge field given by the RR 1-form $C_1$, and $U(n_b)$ gauge fields coming from the 
D8-brane worldvolumes.\footnote{Due to the presence of the singularities at the poles, 
there is also a gauge field given by the reduction of the RR 3-form $C_3$ on $S^2$. However this field is gapped by confinement,
as can be seen from the fact that a D2-brane wrapped on $S^2$ comes with strings attached due to the RR flux on the $S^2$.
A similar mechanism was at work in Type IIB $AdS_6$ duals of 5d SCFTs \cite{Bergman:2018hin}.}
This appears to amount to $U(1)^{s+1} \times \prod_{b=1}^s SU(n_b)$.
However not all of the $U(1)$ vector fields are massless in this background.
This can be seen from the appearance of St\"uckelberg-like terms in the
seven dimensional theory on  $AdS_7$. 

For the RR 1-form such a term comes from the reduction on $M_3$ of Type IIA supergravity
\be 
\label{RRStuck}
S_{C_1} = - \frac{1}{2\,\kappa_{10}^2}\,
\int_{AdS_7\times M_3} dC_5 \wedge H_3 \wedge C_1 =
  \frac{2\,\pi^2}{\kappa_{10}^2}\,N \int_{AdS_7} dC_5\wedge C_1\,.\nonumber  \\ 
\ee
This is a seven-dimensional St\"uckelberg term, where the St\"uckelberg scalar field corresponds to the dual of $C_5$.

There are also St\"uckelberg terms for the D8-brane worldvolume gauge fields that arise from the worldvolume CS terms.
The general form of the worldvolume CS action is given by
\be
S_\text{CS} = T_8\,\int C \wedge \Tr \left(e^{2\pi f + B_2}\right) =T_8\, \int \tilde{C} \wedge \Tr \left(e^{2\pi f}\right) \,,
\ee
where $C$ is the formal sum of all RR potentials, and we have defined the modified RR potentials as
\be
\label{ModifiedRRPotentials}
\tilde{C} \equiv C \wedge e^{B_2} \,,
\ee
such that $d \tilde{C} = \tilde{F}$.
The terms relevant for us are
\be
\label{D8CSTerms}
S_\text{CS} =  T_8\,\int_{AdS_7 \times S^2}\left[ 2\pi \tilde{C}_7 \wedge \Tr f 
+ \frac{1}{2}(2\pi)^2 \tilde{C}_5 \wedge \Tr(f\wedge f)\right] \,.
\ee

The first term gives the St\"uckelberg terms
\be 
\label{D8Stuck1}
S_b^{(1)} = 2\pi\, T_8\, n_b \int_{AdS_7} d\left(\int_{S^2} \tilde{C}_7 \right) \wedge a_b\,,
\ee
where $a_b$ is the diagonal $U(1)$ gauge field of the $b$th D8-brane stack,
and the St\"uckelberg scalar is given by the dual of the reduction of $\tilde{C}_7$ on $S^2$.
These terms imply that the combination $\sum_b n_b a_b$ is massive and therefore not part of the low energy spectrum.

The second term in (\ref{D8CSTerms}) gives the St\"uckelberg terms 
\be
\label{D8Stuck2}
S_b^{(2)} =   (2\pi)^3\, T_8\,\mu_b n_b \int_{AdS_7} d\tilde{C}_5\wedge a_b \,,
\ee
which together with the $C_1$ St\"uckelberg term in (\ref{RRStuck}) seem to imply that 
the combination $NC_1 + \sum_b \mu_b n_b a_b$ is also massive.\footnote{With our conventions, $\kappa_{10}^2=2^6\,\pi^7$, while $T_8=\frac{1}{(2\pi)^8}$. Hence $(2\pi)^3\,T_8=\frac{2\pi^2}{\kappa_{10}^2}$.}
However this not completely correct.
The modified RR potentials defined by (\ref{ModifiedRRPotentials}) are not invariant under $B_2$ gauge transformations,
and so are affected by the large gauge transformations that we used above.
In particular when we cross the plateau $\tilde{C}_7$ undergoes the transformation
\be 
\tilde{C}_7 \rightarrow \tilde{C}_7 + \tilde{C}_5 \wedge N  d\Omega_2 \,,
\ee
giving an additional contribution to (\ref{D8Stuck2}) from (\ref{D8Stuck1}). 
The net effect, given (\ref{D8MagneticFlux}), is to replace $\mu_b$ by $j_b$, 
\be
\label{D8Stuck2Correct}
S_b^{(2)} =   (2\pi)^3\, T_8\, j_b n_b \int_{AdS_7} d\tilde{C}_5\wedge a_b \,.
\ee
So the second massive gauge field is given by $NC_1 + \sum_{b} j_b n_b a_b$.

We are left with $s-1$ massless $U(1)$ gauge fields $A_c$, with $c=1,\ldots, s-1$, that can be parameterized as
\be
\label{MasslessVector}
a_b = \beta_{b}^{(c)} A_c \ , \quad C_1 = \gamma^{(c)} A_c \,,
\ee
where the coefficients $\beta_b^{(c)}$ and $\gamma^{(c)}$ satisfy the conditions
\begin{align}
\label{U(1)Condition1}
&\sum_{b=1}^{s} n_b \beta_b^{(c)} = 0 \, , \\
\label{U(1)Condition2}
&N\gamma^{(c)} + \sum_{b=1}^{s} j_b n_b \beta_b^{(c)} = 0 \,.
\end{align}
We conclude that the gauge symmetry in the $AdS_7$ supergravity background is $U(1)^{s-1}\times \prod_{b=1}^s SU(n_b)$,
in agreement with the global symmetry of the 6d SCFT.

\subsection{Charged states}
\label{sub:probes}

The states charged under the symmetries identified above, and dual to the operators described in the previous section, 
are described by open strings and D0-branes.

\subsubsection{Open strings}

The string-meson ${\cal M}_b$ is dual to an open string between a D8-brane in the $b$th stack and one in the $(b+1)$st stack 
(and located at the origin of $AdS_7$).
Let us refer to this state as $\text{F1}_b$. 
The mass of this string is given by
(with $\alpha'=1$)
\be
\label{StringMass}
M(\text{F1}_{b}) = \frac{1}{2\pi}  \int_{j_b}^{j_{b+1}} dz \sqrt{-\det P[G]_\text{ws}}  =  2(j_{b+1}-j_b)\ ,
\ee
where $P[G]_\text{ws}$ denotes the pullback of the metric in \eqref{eq:10dmetricz} onto the worldsheet, spanning time in $AdS_7$ and $z$. This agrees at large $N$ with the dimension $\Delta_{\mathcal{M}_b} = 2(j_{b+1}-j_b+2)$ of the string-meson operator ${\cal M}_b$ we gave below \eqref{StringMesonGeneral}.
The non-abelian charges also agree, since the open string is charged in the bi-fundamental representation of $SU(n_{b})\times SU(n_{b+1})$.
The $U(1)_c$ charges will be discussed shortly.

We will confirm that this is a BPS state by computing its R-charge.
The charge under $U(1)_R$ is given by the coupling of the string to a fluctuation of the 1-form associated to $U(1)_R$
\be
\delta (\cos\theta \, d\phi) = A_R \,.
\ee 
The relevant terms in the worldsheet action are given by
\be
S_{\text{F1}_b} = \frac{1}{2\pi} \int_{\Sigma_b} B_2 + \int_{\mathbb{R}} (a_{b+1} - a_{b}) \,,
\ee
with $\Sigma_b = [j_b,j_{b+1}]\times \mathbb{R}$.
From (\ref{NSNSGaugePlateau}) we see that the bulk term contributes an amount $N/2$, but only for $b=p$, namely
for the open string in the massless region.
The boundary terms contribute an amount given by $(\mu_{b+1} - \mu_b)/2$.
Adding the two contributions and using (\ref{D8MagneticFlux}) we find for all $b$
\be
Q_R(\text{F1}_{b})= \frac{j_{b+1}-j_b}{2} \,.
\ee
Comparing with the mass in (\ref{StringMass}) we see that this is a BPS state. 
(This is also confirmed by solving the Killing spinor equation -- see appendix \ref{sub:f1prob}.)

\subsubsection{D0-brane (and strings)}

The baryon operators ${\cal B}_j$ are dual to states containing a D0-brane.
The mass of a D0-brane at a point $z$ on the interval (and at the origin of $AdS_7$) is given by 
\be
\label{D0Mass}
M_\text{D0}(z) = e^{-\phi(z)} \sqrt{-G_{00}} 
= \frac{\sqrt{2}}{(9\pi)^2} \left(-\frac \alpha{\ddot \alpha}\right)^{-1/2}(\dot \alpha^2-2 \alpha \ddot \alpha)^{1/2} \,, \label{eq:D0mass}
\ee
with $G$ the ten-dimensional metric in \eqref{eq:10dmetricz}. Generically the D0-brane is located in a massive region with  $F_0(z) \neq 0$.
This requires the presence of $2\pi F_0$ strings between the D0-brane and D8-branes.
From (\ref{RomansMass2}) it follows that if the D0-brane is to the left of the massless region the strings end one at a time on each of the D8-branes to the right up to $z=j_p$,
and if the D0-brane it is to the right of the massless region they do so on each of the D8-branes to the left down to $z=j_{p+1}$.
If the D0-brane happens to be in the massless region there are no strings attached.
The mass of the D0-brane-strings combination is therefore given by
\be
\label{D0StringMass}
M_\text{D0$+$F1}(z) = M_\text{D0}(z)  + \left\{
\begin{array}{ll}
2\sum_{b=a_\text{R}(j)}^p (j_b - z) n_b & \; \text{for} \; j < j_{p} \\[5pt]
0 & \; \text{for} \; j_p < j < j_{p+1} \\[5pt]
2\sum_{b=p+1}^{a_\text{L}(j)} (z-j_b) n_b & \; \text{for} \; j >j_{p+1} \,.
\end{array}
\right.
\ee
The BPS states dual to ${\cal B}_j$ will correspond to minimizing this mass.
While the functional form of $M_\text{D0}(z)$ in (\ref{D0Mass}) looks rather daunting, minimizing the mass of the
D0-brane-strings combination turns out to be quite simple by realizing that the derivative of $M_\text{D0}(z)$ simplifies at the point 
where $\dot\alpha(z)=0$,
\be
\left.\dot{M}_\text{D0}(z)\right|_{\dot{\alpha}=0} = 4\pi F_0(z) \,.
\ee
This minimizes uniquely the mass of the D0-brane-strings combination (\ref{D0StringMass}).
Using the explicit form of the solution (\ref{Solution}) we then find
\begin{align}
\label{MinD0StringMass}
\left. M_\text{D0$+$F1}(z)\right|_{\dot{\alpha}=0} &= 2r_j  + \begin{cases}
2\sum_{b=a_\text{R}(j)}^p (j_b - j) n_b & \; \text{for} \; j < j_{p} \\[5pt]
0 & \; \text{for} \; j_p < j < j_{p+1} \\[5pt]
2\sum_{b=p+1}^{a_\text{L}(j)} (j - j_b) n_b & \; \text{for} \; j > j_{p+1} 
\end{cases}\nonumber \\
&= 2r_\text{max} \,,
\end{align}
where the last equality follows from (\ref{RanksPlateau}).
This agrees with the dimension of the baryon operators ${\cal B}_j$ we gave below \eqref{BaryonPlateau}, namely $\Delta_{{\cal B}_j} = 2r_\text{max}$. 
Importantly this also confirms the conjecture that there is only a single baryon operator in the SCFT,
namely that the $S_N$ symmetry permuting the $N$ baryons ${\cal B}_j$ of the quiver gauge theory is gauged.

The non-abelian charges are trivial since in every stack of $n_b$ D8-branes there is either a string ending on every D8-brane,
giving an $n_b$-fold antisymmetric representation of $SU(n_b)$ which is of course a singlet, or no strings at all. 
The $U(1)_c$ charges will be discussed shortly.

To confirm that this is a BPS state we again compute its R-charge.
The $U(1)_R$ charge will get contributions from the coupling of the D0-brane to the RR 1-form $C_1$,
as well as from the coupling of the string ends to the D8-brane gauge fields.
The contribution of the RR field can be read off from (\ref{RRFluxPlateau}) and is given by
\be 
Q_R(\text{D0}) = \left\{
\begin{array}{ll}
\frac{1}{2}  [jr_{j+1} - (j+1)r_j] & \; \text{for} \; j < j_{p} \\[5pt]
 \frac{1}{2} r_\text{max} & \; \text{for} \; j_p < j < j_{p+1} \\[5pt]
 \frac{1}{2} [(j-N)r_{j+1} - (j-N+1)r_{j}]  &  \; \text{for}\; j > j_{p+1} \,.
 \end{array}
 \right.
 \ee
The contribution of the string ends can be read off from (\ref{D8MagneticFlux}):
\be
Q_R(\text{F1}) = \left\{
\begin{array}{ll}
\frac{1}{2} \sum_{b=a_\text{R}(j)}^p j_b n_b & \; \text{for} \; j < j_{p} \\[5pt]
0 & \; \text{for} \; j_p < j < j_{p+1} \\[5pt]
\frac{1}{2} \sum_{b=p+1}^{a_\text{L}(j)} (j_b - N) n_b & \; \text{for} \; j > j_{p+1} \,.
\end{array}
\right.
\ee
Adding the two contributions and using (\ref{RanksPlateau}) gives
\be
Q_R(\text{D0}+\text{F1}) = \frac{1}{2} r_\text{max} \,,
\ee
confirming that the BPS bound is saturated.

\subsubsection{Matching the $U(1)$ charges}

Recall that there are $s-1$ global $U(1)$ symmetries, $U(1)_c$ with $c=1,\ldots s-1$, parameterized in terms of the RR 1-form and the 
$s$ D8-brane $U(1)$ worldvolume gauge fields by (\ref{MasslessVector}), under the conditions (\ref{U(1)Condition1}) 
and (\ref{U(1)Condition2}).

The charges of the open string $\text{F1}_b$ are simply given by
\be\label{eq:F1bb}
Q_c(\text{F1}_b) = \beta_{b+1}^{(c)} - \beta_{b}^{(c)} \,.
\ee
These can be matched to the string-meson charges (\ref{StringMesonCharges}), yielding $(s-1)^2$ conditions, 
which together with the $s-1$ conditions in (\ref{U(1)Condition1}), can be solved for the $s(s-1)$ variables $\beta_b^{(c)}$.
This is a rather cumbersome calculation, the outline of which we sketch in appendix \ref{app:charges}.
Of course this by itself does not provide a test of the duality.

Now let us consider the D0-brane.
There are in general three contributions to the $U(1)_c$ charge of this state.
The first is the RR charge of the D0-brane, which contributes an amount $\gamma^{(c)}$.
The second is the D8-brane worldvolume charge at the endpoint of the D0-D8 string,
which contributes an amount $\beta_b^{(c)}$ for each string ending on a D8-brane in the $b$th stack.
The third is a D8-brane worldvolume charge induced by the RR flux sourced by the D0-brane,
which contributes $\frac{1}{2}\beta_b^{(c)}$ for each D8-brane located to the right of the D0-brane,
and $-\frac{1}{2}\beta_b^{(c)}$ for each D8-brane located to the left of the D0-brane.\footnote{This is due to
the D8-brane worldvolume coupling $\int C_7 \wedge f_b = \int F_8 \wedge a_b$. The D8-brane captures half of
the total $F_8$ flux sourced by the D0-brane, and the sign depends on the relative orientation of the D8-brane and the D0-brane.
In fact the string creation effect can be understood from the 
requirement that the worldvolume charge remains invariant as the D0-brane crosses the D8-brane.}
The sum of the last two contributions is independent of the position of the D0-brane.
The total charge is given by 
\begin{align}
\label{eq:D0F1U1charge}
Q_c(\text{D0}+\text{F1}) &= \gamma^{(c)}  + \frac{1}{2} \left[ \sum_{b={a_\text{R}(j)}}^s n_b \beta_b^{(c)}  - \sum_{b=1}^{a_\text{L}(j)} n_b \beta_b^{(c)} \right] + \nonumber \\
& \ \ \ \, +\begin{cases}
- \sum_{b=a_\text{R}(j)}^p n_b \beta_b^{(c)} & \; \text{for} \; j \leq j_{p} \\[5pt]
0 & \; \text{for} \; j_p < j < j_{p+1} \\[5pt]
\sum_{b=p+1}^{a_\text{L}(j)} n_b \beta_b^{(c)} & \; \text{for} \; j \geq j_{p+1}
\end{cases} \nonumber \\
&=  \gamma^{(c)}  + \frac{1}{2} \left[ \sum_{b=p+1}^s n_b \beta_b^{(c)}  - \sum_{b=1}^{p} n_b \beta_b^{(c)} \right]  \nonumber \\
&=  \gamma^{(c)}  -  \sum_{b=1}^{p} n_b \beta_b^{(c)} \,,
\end{align}
where in the second equality we used the fact that the charge is independent of the position of the D0-brane to set $j_p<j<j_{p+1}$,
and in the third equality we used (\ref{U(1)Condition1}).

Now comes a non-trivial test of the duality.
Given the values of $\beta_b^{(c)}$ obtained by matching the string-meson charges
and the condition (\ref{U(1)Condition2}) we obtain $\gamma^{(c)}$.
Then the above charges should be compared with the baryon charges in (\ref{BaryonChargesPlateau}).
Doing this in the general case is cumbersome.
We will do it here for $s=2$,\footnote{In this case $c=1$, so we can drop this index from $\beta_b$ and $\gamma$.} and later for some specific examples. (See appendix \ref{app:charges} for the general expressions as well as the $s=3$ case. However we have not proven the relation $Q_c(\text{D0}+\text{F1})=Q_c(\mathcal{B}_j)$ for arbitrary $s$.)

In this case condition (\ref{U(1)Condition1}) reduces to
\be
n_1\beta_1 + n_2\beta_2 = 0
\ee
and then condition (\ref{U(1)Condition2}) reduces to
\be
N\gamma = n_1(j_2-j_1)\beta_1\,.
\ee
In addition, (\ref{MaxRankPlateau}) reduces to
\be 
r_\text{max} = j_1n_1 = (N-j_2)n_2
\ee
so 
\be 
j_1 n_1 + j_2 n_2 = Nn_2 \,.
\ee
Matching the string-meson charge (\ref{StringMesonCharges}) reduces to
\be
\beta_2 - \beta_1 = \frac{n_1+n_2}{r_\text{max}} + \frac{n_1n_2(j_2-j_1)}{r_\text{max}^2} 
= \frac{n_1 + n_2}{n_1 j_1} + \frac{n_1n_2(j_2-j_1)}{n_1^2 j_1^2} \,.
\ee
We then find
\be 
\beta_1 = - \frac{n_2(j_1n_1 + j_2n_2)}{n_1 j_1^2(n_1 + n_2)} 
\ee
and thus
\be
\gamma =- \frac{n_2^2(j_2 - j_1)}{j_1^2(n_1 + n_2)} \,.
\ee
The charge of the D0-brane-string combination is then given by
\be
Q_1(\text{D0}+\text{F1}) = \gamma  -  n_1 \beta_1 = \frac{n_2}{j_1} = \frac{n_1 n_2}{r_\text{max}} \,,
\ee
in precise agreement with the charge of the dual baryon operator (\ref{BaryonChargesPlateau}).

\subsubsection{Holographic chiral-ring relation}

The chiral-ring relation (\ref{ChiralRingPlateau}) also has a nice geometrical description.
Consider a Euclidean D2-brane that wraps the entire internal space $M_3$.
The coupling of the background fluxes to the D2-brane worldvolume induces tadpoles 
which must be cancelled by attaching both D0-branes and strings to the D2-brane.

The NSNS flux on $M_3$ induces a worldvolume magnetic charge via the DBI action, whose expansion contains the term 
$\int_{M_3} B_2 \wedge *f = \int_{M_3} \varphi \, H_3$, where $\varphi$ is the D2-brane worldvolume magnetic-dual scalar potential.
The amount of magnetic charge is given by the NSNS flux on $M_3$, namely $N$.
This is a pointlike charge in the compact three-dimensional Euclidean worldvolume of the D2-brane, and constitutes a tadpole which must be cancelled by 
having $N$ D0-brane worldlines end on the D2-brane at a point.

The RR flux on $S^2$ induces a worldvolume electric charge via the worldvolume CS coupling 
$\int_{M_3} C_1\wedge f = \int_{M_3} \tilde{F}_2 \wedge a$, where $a$ is the D2-brane worldvolume gauge potential.
The amount of charge in a given segment $j_b < z < j_{b+1}$ is given by the RR flux on $S^2$ in that segment, i.e. (\ref{RRFluxPlateau}).
This charge also constitutes a tadpole, which must be cancelled by open string worldsheets ending
on the D2-brane along the above segment, where the number of open strings $\text{F1}_{b}$ is given by the electric charge in the 
corresponding segment.
This gives
\be
N_{\text{F1}_b} = 
\begin{cases}
\sum_{a=1}^{b} j_a n_a &   \text{for}\; b< p \\
r_\text{max} &   \text{for}\; b=p  \\
\sum_{a=b+1}^s (N-j_a) n_a   &  \text{for}\; b>p \,.
  \end{cases}
\ee

\begin{figure}[h!]
\center
\includegraphics[width=0.35\textwidth]{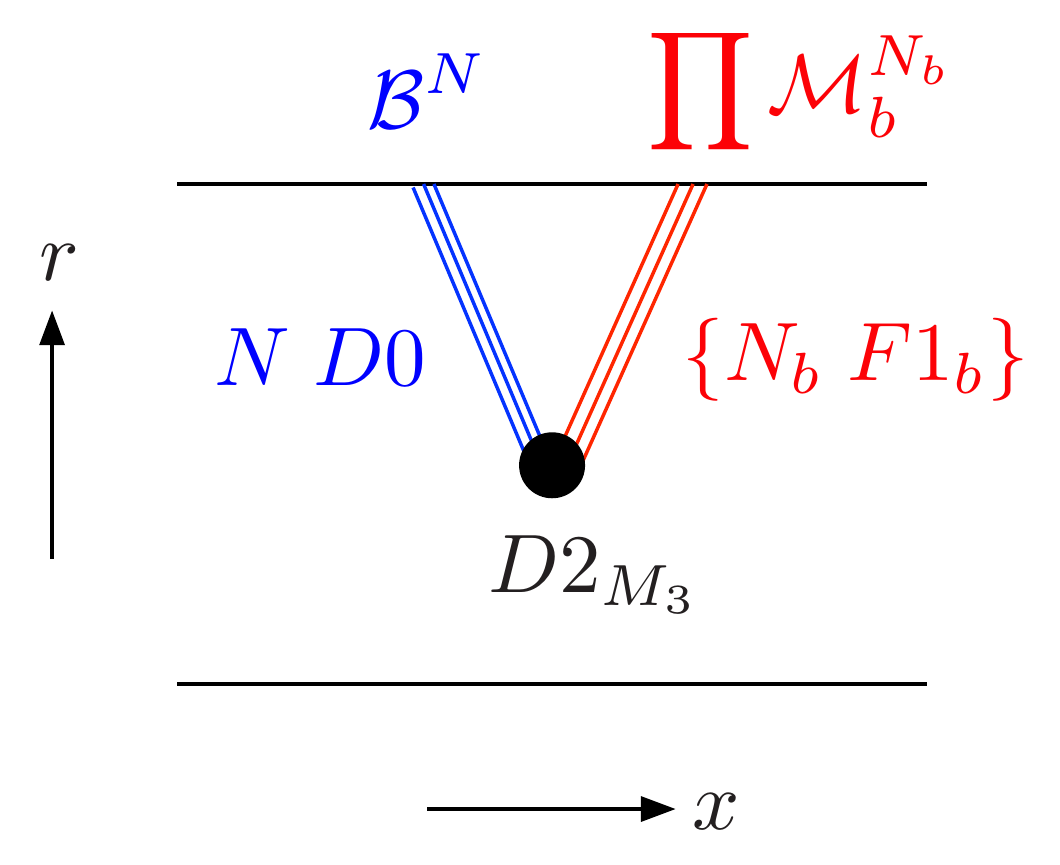} 
\caption{Holographic description of the chiral-ring relation. The vertical axis (labeled by $r$) represents the radial coordinate of $AdS_7$, with $r=0$ at the bottom and $r=\infty$  at the top; the horizontal axis (labeled by $x$) represents any of the other coordinates.}
\label{ChiralRingHolographic}
\end{figure}

The wrapped Euclidean D2-brane therefore describes a process in which $N$ D0-branes turn into a collection of open strings, 
with $N_{\text{F1}_{b}}$ strings between the $b$th and $(b+1)$st D8-brane stack, or vice versa; see Fig.~\ref{ChiralRingHolographic}.
This is precisely the content of the chiral-ring relation in (\ref{ChiralRingPlateau}).
The number of open strings $N_{\text{F1}_{b}}$ appears as the total power of the string-meson ${\cal M}_b$ on the RHS.
The neutral meson factors, which we omitted, account for the deficit or excess of energy in the process.

\section{Plateau-less cases}
\label{PlateauLess}

For a linear quiver without a plateau the rank is maximized at the edge.
We will assume this to be the right edge, and therefore $r_\text{max} = r_{N-1}$.
Imposing $r_0=r_N=0$ in this case we find
\be
\label{RanksNoPlateau}
r_j = (N-j) r_\text{max} - \sum_{b=a_\text{R}(j)}^s (j_b - j) n_b \,,
\ee
and 
\be
\label{MaxRankNoPlateau}
r_\text{max} = r_{N-1} =  \frac{1}{N}\sum_{a=1}^s j_a n_a \,.
\ee
The expression for the non-anomalous $U(1)$'s is the same as before (\ref{AnomalyFreeSymmetries}).
The expressions for the string-meson operators and their charges are also the same.
The baryon operators are slightly different in this case.
There are again $N$ baryons, now given by
\be
\label{BaryonNoPlateau}
{\cal B}_j =
(x_j)^{r_j} \prod_{b=a_\text{R}(j)}^{s-1} \left[\left(\prod_{i=j+1}^{j_b-1} \tilde{x}_i\right)\cdot y_b\right]^{n_b}
 \left[\left(\prod_{i=j+1}^{N-2} \tilde{x}_i\right)\cdot y_s\right]^{n_s - r_\text{max}} \; \text{for} \; 0\leq j \leq  N-2 \,, \nonumber \\
\ee
and
\be
\label{BaryonNoPlateau2}
{\cal B}_{N-1} = (\tilde{y}_s)^{r_\text{max}} \,.
\ee
All of these have a scaling dimension $\Delta_{{\cal B}_j} = 2r_\text{max}$, they all transform under $SU(n_s)$ in the $(n_s - r_\text{max})$-fold antisymmetric representation, and they all carry only $U(1)_{s-1}$ charge:
\be
\label{BaryonChargesNoPlateau}
Q_a({\cal B}_j) = \left\{
\begin{array}{ll}
n_{s-1} & \; \text{for} \; a=s-1 \\
0 & \; \text{for} \; a\neq s-1 \,.
\end{array}
\right.
\ee
The chiral-ring relation is also a little different in the plateau-less case, and is given, modulo neutral meson factors, by
\be
\label{ChiralRingNoPlateau}
\prod_{j=0}^{N-1} {\cal B}_j \sim 
\prod_{a=1}^{s-1} ({\cal M}_a \cdots {\cal M}_{s-1})^{n_a j_a} \,.
\ee
In this case both sides transform in the $N(n_s-r_\text{max})$-fold antisymmetric representation of $SU(n_s)$.

The supergravity solutions dual to the plateau-less quivers 
do not have a massless region, 
and the Romans mass for $j<z<j+1$ is given by
\be
\label{RomansMassNoPlateau}
2\pi F_0(z) =  - r_\text{max} + \sum_{b=a_\text{R}(j)}^s n_b \,.
\ee
A convenient gauge choice for the NSNS field in this case is
\be
\label{NSNSGaugeNoPlateau}
B_2 = 
\begin{cases}
h(z) \, d\Omega_2 & \text{for}\;  z < N-1 \\
  (\pi N + h(z))\, d\Omega_2 & \text{for}\; z =N \,, \\
\end{cases} 
\ee
namely the large gauge transformations are performed in the very last segment beyond the last stack of D8-branes.
The D6-brane Page charge in the segment $j < z < j+1$  is consequently given by
\be
\label{RRFluxNoPlateau}
\int_{S^2} \frac{\tilde{F}_2}{2\pi}  = 
\begin{cases}
 - \sum_{b=1}^{a_\text{R}(j)-1} j_b n_b   &   \text{for}\;  j < j_s \\
-  N r_\text{max} = - N r_{N-1} &   \text{for}\;  j = j_s  \,,
  \end{cases}
\ee
and the associated D8-brane worldvolume magnetic fluxes are 
\be
\int_{S^2} \frac{f_b}{2\pi} = \mu_b = j_b \,. 
\ee

The analysis of the symmetries is unchanged, as is the analysis of the open strings dual to the string-meson operators.

The results for the D0-brane are slightly different.
The D0-brane has $2\pi F_0$ strings attached, where $F_0$ is given by (\ref{RomansMassNoPlateau}).
It follows from (\ref{RomansMassNoPlateau}) that for $j_s -1 < z < j_s$ there are $n_s - r_\text{max}$ strings,
and $n_b$ additional strings for each stack of D8-branes that the D0-brane crosses as it moves to the left.
Therefore, unlike in the solutions containing a massless region, this state is charged under the last flavor symmetry $SU(n_s)$,
transforming in the $(n_s- r_\text{max})$-fold antisymmetric representation, in agreement with the dual baryon operators.

The $U(1)_c$ charges are given by
\begin{align}
Q_c(\text{D0}+\text{F1}) &= \gamma^{(c)} + \frac{1}{2} \left[ \sum_{b={a_\text{R}(j)}}^s n_b \beta_b^{(c)}  - \sum_{b=1}^{a_\text{L}(j)} n_b \beta_b^{(c)} \right] 
 \nonumber \\
& \ \ \ \,+ \left[\sum_{b=a_\text{R}(j)}^{s-1} n_b \beta_b^{(c)} + (n_s - r_\text{max}) \beta_s^{(c)}\right] \,,
\end{align}
where the first term is the contribution of the D0-brane RR charge, the second term is the contribution of the induced D8-brane
worldvolume charges, and the third term is the contribution of the string end charges.
Establishing that these agree with the $U(1)_c$ charges of the baryon operators is again a tedious, yet straightforward exercise.
We will do it for a specific example with $s=2$ below.

The mass of the D0-brane-strings combination is given by 
\be
\label{D0StringMassNoPlateau}
M_\text{D0$+$F1}(z) = M_\text{D0}(z)  + 
2\sum_{b=a_\text{R}(j)}^{s-1} (j_b - z) n_b + 2(n_s - r_\text{max})(j_s - z) \,.
\ee
This again has a unique minimum when $\dot{\alpha}=0$, in which case we again find that 
\be 
M(\text{D0} + \text{F1}) = 2r_\text{max} \,,
\ee
in agreement with the dimension of the baryon operators, and confirming again that the baryon permutation symmetry is gauged in the field theory.

The geometric description of the chiral-ring relation (\ref{ChiralRingNoPlateau}) is similar to the case with a massless region.
A Euclidean D2-brane wrapping $M_3$ contains $N$ units of magnetic charge induced by the NSNS flux on $M_3$,
and $\frac{\tilde{F}_2}{2\pi}$ units of electric charge induced by the RR flux on $S^2$ in a given segment.
As before, these charges must be canceled by having $N$ D0-brane worldlines and $N_{\text{F1}_{b}}$ string worldsheets between $z=j_b$ and $z=j_{b+1}$
end on the D2-brane, where in this case
\be
 N_{\text{F1}_{b}} = \sum_{a=1}^{b} j_a n_a \,.
\ee
This is precisely the content of the chiral-ring relation (\ref{ChiralRingNoPlateau}).

\section{Examples}
\label{Examples}

\subsection{\texorpdfstring{A uniform quiver: M5-branes on $\mathbb{C}^2/\mathbb{Z}_k$}{A uniform quiver: M5-branes on C2/Zk}}
\label{sub:M5}

\begin{figure}[ht!]
\centering
\includegraphics[scale=.65]{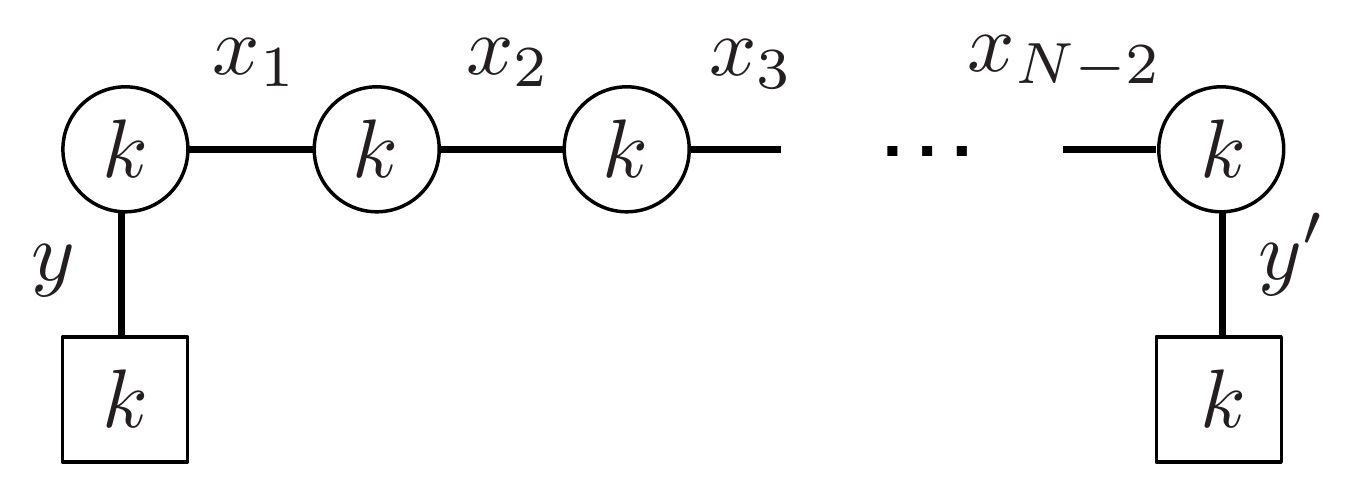} 
\caption{Uniform quiver: $N$ M5-branes on $\mathbb{C}^2/\mathbb{Z}_k$ at a generic point on the tensor branch.}
\label{QuiverUniform}
\end{figure}

The simplest class of examples corresponds to $N$ M5-branes on $\mathbb{C}^2/\mathbb{Z}_k$, which at a generic point on the tensor branch
is given by the uniform linear quiver gauge theory shown in Fig.~\ref{QuiverUniform}.
The global symmetry is $SU(k)\times SU(k)\times U(1)$, where the single anomaly-free $U(1)$ current is given by
\be 
J^\mu = I_y^\mu + \sum_{i=1}^{N-2} J^\mu_i - I_{y'}^\mu \,.
\ee

The charged operators consist of a single string-meson
\be
\label{StringMeson1}
{\cal M} = y \cdot  \left(\prod_{i=1}^{N-2}  x_i \right) \cdot \tilde{y}' \,,
\ee
and $N$ baryons
\be
\label{Baryon1}
{\cal B}_0 = \det\, y \,, \; {\cal B}_i  = \det\, x_i \,, \; {\cal B}_{N-1}  = \det\, \tilde{y}' \,,
\ee
whose properties are summarized in Table~\ref{tab:BMcharges1}.
The chiral-ring relation takes the simple form
\be
\label{ChiralRing1}
\prod_{i=0}^{N-1} {\cal B}_i \propto \det\, {\cal M} \ .
\ee

\begin{table}[h!]
\begin{center}
\begin{tabular}{c|c|c|c}
 operator & dimension & $Q$ & $SU(k)\times SU(k)$ \\
 \hline \hline
 ${\cal B}$ & $2k$ & $k$ & $({\bf 1},{\bf 1})$ \\
 \hline
 ${\cal M}$ & $2N$ & $N$ & $({\bf k},\bar{\bf k})$ \\
 \end{tabular}
 \end{center}
\caption{Spectrum of charged BPS operators in the uniform quiver.}
\label{tab:BMcharges1}
\end{table}

The dual supergravity background has $F_0=0$ everywhere, and the solution is given by \cite{Apruzzi:2013yva}\footnote{The notation of \cite{Apruzzi:2013yva}
is slightly different. 
See \cite[App. A]{Apruzzi:2017nck} for the dictionary between the notation used there and the one adopted here (i.e. the $z$ coordinate of \cite{Cremonesi:2015bld}).}
\be
\label{UniformQuiverSolution}
\alpha(z) = \frac{(9\pi)^2}{2} k z(N - z) \,.
\ee
Note that this is invariant under $z\rightarrow N-z$, which is consistent with the reflection symmetry of the quiver.
The singularities at the poles $z=0$ and $z=N$ correspond to the semi-infinite D6-branes, or equivalently to two stacks of $k$
D8-branes at $z=1$ and at $z=N-1$, with worldvolume magnetic fluxes $\mu_1=1$ and $\mu_2=-1$, respectively.

\begin{figure}[ht!]
\centering
\includegraphics[scale=0.4]{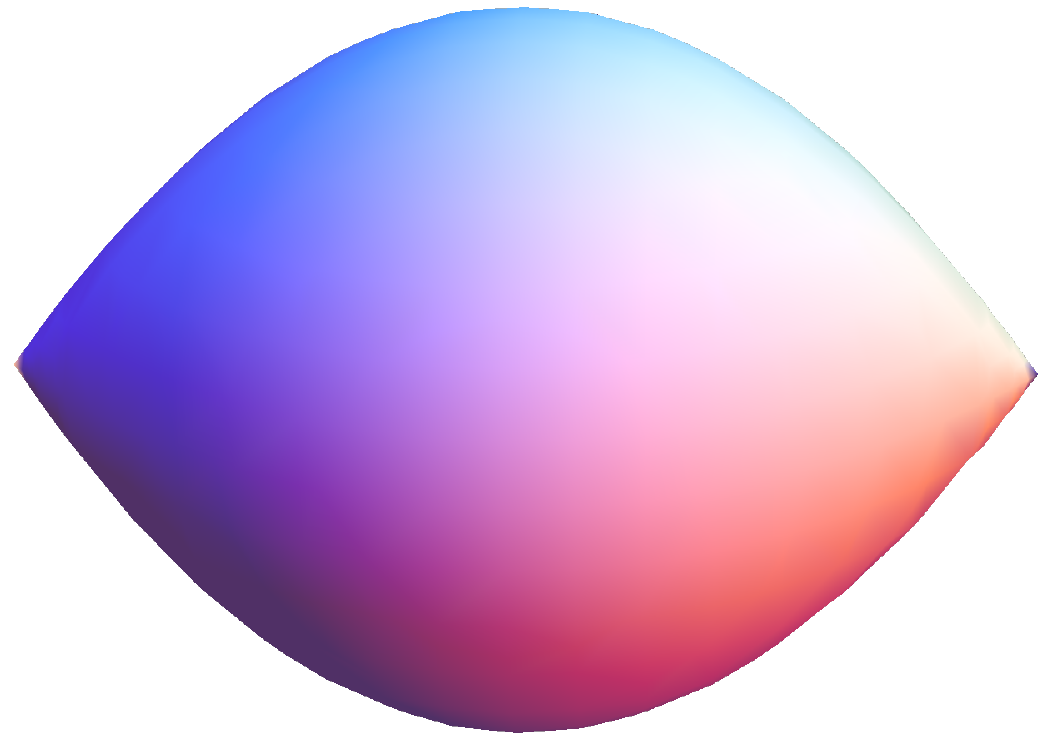} 
\caption{A two-dimensional projection of the internal space $M_3$ dual to the uniform quiver.}
\label{massless}
\end{figure}
The RR flux on $S^2$ is independent of $z$ and given by 
\be
\int_{S^2} \frac{\tilde{F}_2}{2\pi} =  -  k \,.
\ee
The $SU(k)\times SU(k)$ symmetry is realized as the non-abelian part of the D8-brane worlvolume gauge symmetry.
There is also a single massless $U(1)$ gauge field $A$ parameterized by the two diagonal D8-brane worlvolume $U(1)$ gauge fields and the RR gauge field as
\be
a_1 = \beta_{1} A \ , \quad a_2 = \beta_{2} A \ , \quad C_1 = \gamma A \ ,
\ee
where
\be
\label{MasslessVectorUniform}
(\beta_1,\beta_2,\gamma) = \left(-\frac{N}{2},\frac{N}{2},k\left(1-\frac{N}{2}\right)\right) \,.
\ee

The string-meson operator ${\cal M}$ is dual to an open string connecting the two poles, or equivalently between the two D8-brane stacks.
This transforms in the bi-fundamental representation of $SU(k)\times SU(k)$, and 
given (\ref{MasslessVectorUniform}), it carries $N$ units of $U(1)$ charge, in agreement with the dual operator.
The mass is given by (\ref{StringMass}), which in this case is
\be
M(\text{F1}) = 2(N-2) \,,
\ee
which agrees for large $N$ with the operator dimension.
Note that the mass of a string between the two poles is $2N$, in precise agreement with the operator dimension.
This is consistent with the large $N$ equivalence of the two descriptions of the sources at the poles.

The baryon operator is dual to the D0-brane.
This is a singlet of $SU(k)\times SU(k)$, and its $U(1)$ charge is given by
\be
Q(\text{D0}) = \gamma +  \frac{1}{2} (  n_2 \beta_2  - n_1 \beta_1 ) = k\left(1-\frac{N}{2}\right) + \frac{kN}{2} = k \,,
\ee
in agreement with the charge of the baryon operator.
The BPS condition $\dot{\alpha}=0$ has a unique solution that fixes the position of the D0-brane at $z_\text{min}=N/2$,
and its mass is
\be
M(\text{D0}) = 2k \,,
\ee 
in agreement with the dimension of the dual baryon operator.

The geometric description of the chiral-ring relation (\ref{ChiralRing1}) reduces in this example
to an instantonic D2-brane mediated process turning $N$ D0-branes into $k$ strings.

\subsection{A uniformly rising quiver}
\label{sub:RisingQuiver}

\begin{figure}[ht!]
\centering
\includegraphics[scale=.6]{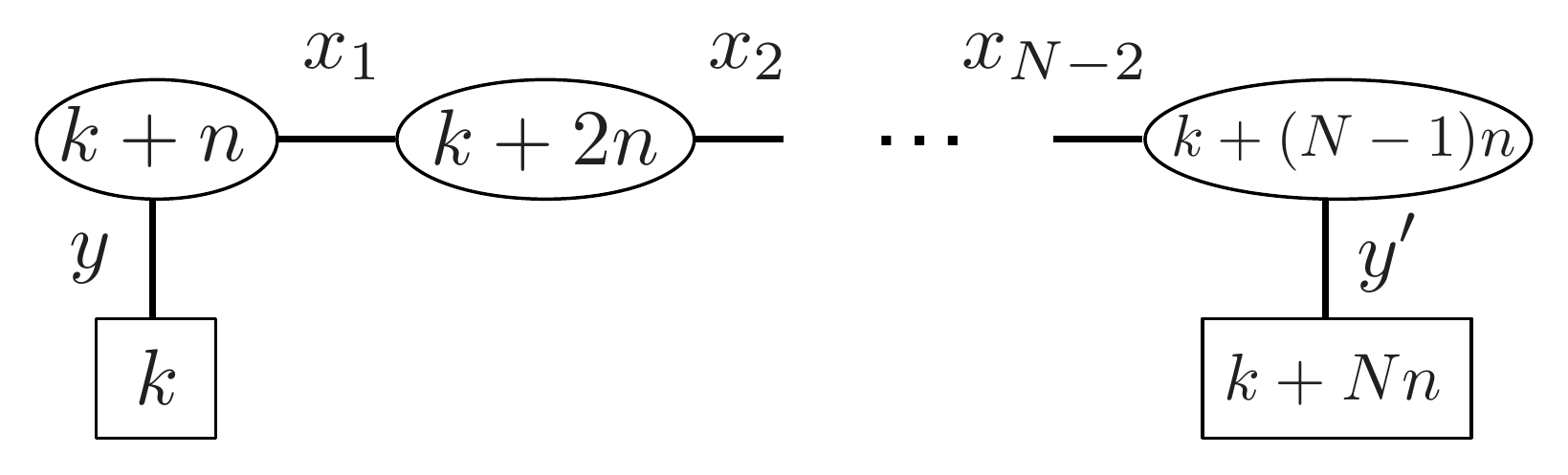} 
\caption{A uniformly rising quiver.}
\label{QuiverRising}
\end{figure}

As our second example we will take a quiver with a uniformly increasing rank, Fig.~\ref{QuiverRising}.
This is an example of a plateau-less quiver.
For $n=0$ this reduces to the previous example of the uniform quiver.
The global symmetry is $SU(k)\times SU(k+Nn) \times U(1)$, except if $k=0$ in which case it is just $SU(Nn)$.
The anomaly-free $U(1)$ current is now given by
\be 
\label{AnomalyFreeCurrent2}
J^\mu = \frac{k+nN}{k+n} I_y^\mu + \sum_{j=1}^{N-2} \frac{k(k+nN)}{(k+jn)(k+(j+1)n)} \, J^\mu_i - \frac{k}{(k+(N-1)n)} I_{y'}^\mu \,. \nonumber \\
\ee
For $k=0$ the first term is absent and there is no anomaly-free $U(1)$ current.

The charged BPS operators include a string-meson of the same form as (\ref{StringMeson1}), now
transforming as a bi-fundamental of $SU(k)\times SU(k+Nn)$, 
and $N$ baryons
\be
\label{Baryon2}
{\cal B}_0 &=& (y)^k \cdot  \left[ \left(\prod_{j=1}^{N-2} \tilde{x}_j \right) {y}'\right]^n \\
{\cal B}_i &=& (x_i)^{k+in} \cdot \left[ \left(\prod_{j=i+1}^{N-2} \tilde{x}_j \right) {y}'\right]^n \\
{\cal B}_{N-1} &=& (\tilde{y}')^{k+(N-1)n}
\ee
which transform in the $n$-fold antisymmetric representation of $SU(k+Nn)$.
The properties of these operators are summarized in Table~\ref{tab:BMcharges2}.
The chiral-ring relation is given by
\be
\label{ChiralRing2}
\prod_{i=0}^{N-1} {\cal B}_i \sim  ({\cal M})^k \,,
\ee
where both sides transform in the $Nn$-fold antisymmetric representation of $SU(k+Nn)$.

\begin{table}[h!]
\begin{center}
\begin{tabular}{c|c|c|c}
 operator & dimension & $Q$ & $SU(k)\times SU(k+Nn)$ \\
 \hline \hline
 ${\cal B}$ & $2[k+(N-1)n]$ & $k$ & $({\bf 1},[{\bf k+Nn}]^n_\text{asym})$ \\[5pt]
 ${\cal M}$ & $2N$ & $N$ & $({\bf k},\overline{\bf k+Nn})$ \\
 \end{tabular}
 \end{center}
\caption{Local operators and their charges.}
\label{tab:BMcharges2}
\end{table}

For $k=0$ we lose the string-meson ${\cal M}$ and the baryon ${\cal B}_0$.
On the other hand there is a new independent baryon operator which we can call ${\cal B}_0$,
\be
{\cal B}_0 = \left[ \left(\prod_{i=1}^{N-2} \tilde{x}_i \right) \tilde{y}'\right]^n_\text{asym}\,,
\ee
with all the same properties.
The chiral-ring-like relation reduces in this case to
\be
\prod_{i=0}^{N-1} {\cal B}_i \sim 1 \,. 
\ee

The solution dual to this theory has $2\pi F_0= n$ and 
\be
\label{SolutionRising}
\alpha(z) = \frac{(9\pi)^2}{6}   z (N-z)[3k+n(N+z)] \,.
\ee
This reduces to the solution for the uniform quiver (\ref{UniformQuiverSolution}) for $n=0$.
The poles $z=0$ and $z=N$ are generically singular, and correspond to $k$ D8-branes with $\mu_1=1$ unit of magnetic flux at $z=1$ 
and $k+Nn$ D8-branes with $\mu_2=N-1$ units of magnetic flux at $z=N-1$.
For $k=0$ there is no singularity at $z=0$.

\begin{figure}[ht!]
\centering
\includegraphics[scale=0.6,angle=0,origin=c]{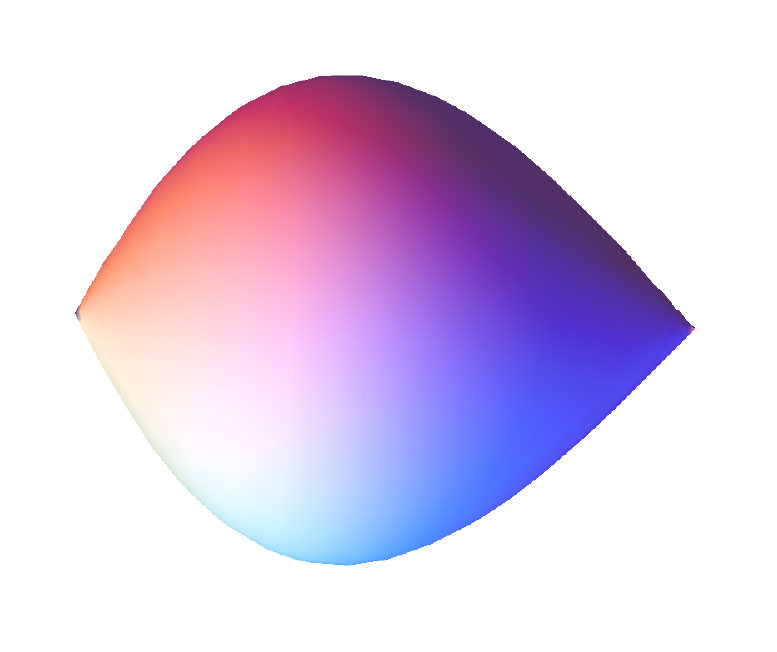} 
\caption{A two-dimensional projection of the internal space $M_3$ dual to the uniformly rising quiver.}
\label{lemon}
\end{figure}

The RR flux in this case is also independent of $z$ and given by 
\be
\int_{S^2} \frac{\tilde{F}_2}{2\pi} = - k \,.
\ee
The D8-branes realize the $SU(k)\times SU(k+Nn)$ part of the gauge symmetry, and for $k\neq 0$ there is again a single massless $U(1)$ gauge field
given in this case by 
\be
\label{MasslessVectorRising}
(\beta_1,\beta_2,\gamma) = \frac{1}{2k+Nn} \left(-N(k+Nn),kN,k(2-N)(k+Nn)\right) \,.
\ee
Note that this reduces to the parameterization of the massless $U(1)$ of the uniform quiver (\ref{MasslessVectorUniform}) for $n=0$.

The string-meson operator is again dual to an open string between the two D8-brane stacks, 
and has the same mass as before.
This state is charged in the bi-fundamental representation of $SU(k)\times SU(k+Nn)$, and carries
$N$ units of $U(1)$ charge, as is easily read off from (\ref{MasslessVectorRising}).
For $k=0$ this state is absent.

The baryon operator is dual to a D0-brane, which now has $n$ strings connecting it to the south pole, or equivalently to $n$ of the $k+Nn$ 
D8-branes at $z=N-1$. 
Since the massless spectrum of the D0-D8 string consists of a single fermion, this state transforms in the $n$-fold antisymmetric
representation of $SU(k+Nn)$, in agreement with the baryon operator.
The $U(1)$ charge of this state is given by 
\be
Q(\text{D0}+\text{F1}) =  \gamma + \frac{1}{2}[(k+nN)\beta_2 - k\beta_1]  - n\beta_{2} = k \,,
\ee
also in agreement with the dual operator.
The minimal mass of the D0-brane-strings combination is given by 
\be
M(\text{D0}+\text{F1}) = 2r_\text{max} = 2(k+n(N-1)) \,,
\ee
in agreement with the dimension of the baryon operator.

The geometric description of chiral-ring relation (\ref{ChiralRing2}) again reduces to the identification of $N$ D0-branes with
$k$ open strings via a wrapped Euclidean D2-brane.


\subsection{A simple symmetric quiver} 
\label{sub:symflav}

\begin{figure}[ht!]
\centering
\includegraphics[scale=.75]{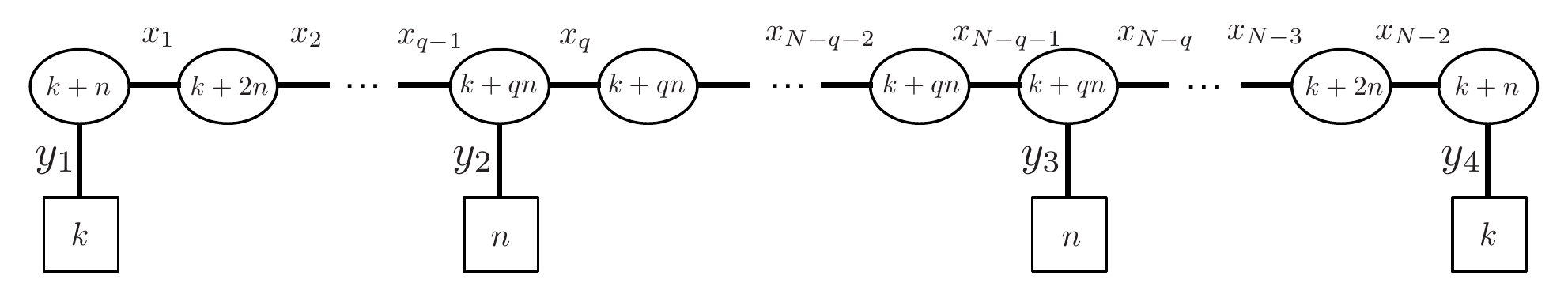} 
\caption{The symmetric quiver.}
\label{QuiverSymmetric}
\end{figure}

Our third and final example is the quiver shown in Fig.~\ref{QuiverSymmetric}.
For $n=0$ this reduces again to the uniform quiver.
For $q=1$ it also reduces to the uniform quiver with $k$ replaced by $k+n$.
More generally this theory can be viewed as a Higgs branch deformation of the uniform quiver with $k$ replaced by $k+qn$ 
\cite{Gaiotto:2014lca}.
The global symmetry is $SU(k)^2\times SU(n)^2\times U(1)^3$, where the three $U(1)$ currents are given by
\be
\label{ThreeU(1)s}
J_1^\mu &=& \frac{n}{k+n} I_1^\mu + \sum_{i=1}^{q-1} \frac{nk}{(k+(i+1)n)(k+in)} \, J^\mu_i  -  \frac{k}{k+qn} I^\mu_2 \nonumber \\ 
J_2^\mu &=& \frac{n}{k+qn}I_2^\mu + \frac{n^2}{(k+qn)^2} \sum_{i=q}^{N-q-1}  J_i^\mu - \frac{n}{k+qn} I_{3}^\mu \\
J_3^\mu &=& \frac{k}{k+qn} I_3^\mu + \sum_{i=N-q}^{N-2}  \frac{nk}{(k+(N-i)n)(k+(N-i-1)n)}\, J^\mu_i - \frac{n}{k+n} I_{4}^\mu \,. \nonumber
\ee
For the special case of $k=0$ there is only the second $U(1)$ symmetry.
The spectrum of charged BPS operators includes three string-mesons
\be
\label{StringMesons}
{\cal M}_1 &=& y_1 \cdot \left(\prod_{i=1}^{q-1} x_i \right) \cdot \tilde{y}_2 \\
{\cal M}_2 &=& y_2 \cdot \left(\prod_{i=q}^{N-q-1} x_i\right) \cdot \tilde{y}_3 \ ,\\
{\cal M}_3 &=& y_3 \cdot \left(\prod_{i=N-q}^{N-2} x_i\right) \cdot \tilde{y}_4
\ee
and $N$ baryons 
\be
\label{BaryonLike}
{\cal B}_0 &=& (y_1)^k \cdot  \left[ \left(\prod_{j=1}^{q-1} \tilde{x}_j \right) y_2\right]^n \\
{\cal B}_i &=& 
\begin{cases}
(x_i)^{k+in} \cdot \left[ \left(\prod_{j=i+1}^{q-1} \tilde{x}_j \right) y_2 \right]^n & \text{for}\; 1\leq i \leq q-1 \\
(x_i)^{k+qn} = \det \, x_i & \text{for}\; q\leq i \leq  N-q-1 \\
(x_i)^{k+(N-i-1)n} \cdot      \left[\tilde{y}_3 \left(\prod_{j=N-q}^{i-1} \tilde{x}_j \right) \right]^n  & \text{for}\; N-q \leq i \leq N-2 \\
\end{cases} \\
{\cal B}_{N-1} &=& (\tilde{y}_4)^k \cdot \left[\tilde{y}_3 \left(\prod_{j=N-q}^{N-2} \tilde{x}_j \right) \right]^n \,,
\ee
whose properties are summarized in Table~\ref{tab:BMcharges}.
The chiral-ring relation is given by
\begin{equation}
\label{ChiralRing3}
\prod_{i=0}^{N-1} {\cal B}_i \sim ({\cal M}_1)^k \cdot ({\cal M}_2)^{k+nq} \cdot ({\cal M}_3)^k =
\det ({\cal M}_1 \cdot {\cal M}_2 \cdot {\cal M}_3)
(\det {\cal M}_2)^q \,.
\end{equation}

\begin{table}[h!]
\begin{center}
\begin{tabular}{c|c|c|c|c|c}
 operator & dimension & $Q_1$ & $Q_2$ & $Q_3$ & $SU(k)^2\times SU(n)^2$ \\
 \hline \hline
 ${\cal B}$ & $2(k+nq)$ & 0 & $\frac{n^2}{k+nq}$ & 0 & $({\bf 1},{\bf 1},{\bf 1},{\bf 1})$ \\
 \hline
 ${\cal M}_1$ & $2(q+1)$ & $1$ & $-\frac{n}{k+nq}$ & 0 & $({\bf k},{\bf 1},\bar{\bf n},{\bf 1})$ \\
  \hline
 ${\cal M}_2$ & $2(N-2q+2)$ & $-\frac{k}{k+nq}$ & $\frac{n(nN+2k)}{(k+nq)^2}$ & $-\frac{k}{k+nq}$ & $({\bf 1},{\bf 1},{\bf n},\bar{\bf n})$ \\
\hline
 ${\cal M}_3$ & $2(q+1)$ & $0$ & $-\frac{n}{k+nq}$ & $1$ & $({\bf 1},\bar{\bf k},{\bf 1},{\bf n})$ \\
 \end{tabular}
 \end{center}
\caption{Spectrum of charged BPS operators in the symmetric quiver}
\label{tab:BMcharges}
\end{table}
For the case $k=0$ we lose the string-mesons ${\cal M}_1, {\cal M}_3$, and the baryons ${\cal B}_0, {\cal B}_{N-1}$,
and gain two new baryon-like operators given by
\be
{\cal B}_0 &=& \left[ \left(\prod_{i=1}^{q-1} \tilde{x}_i \right) y_2 \right]^n_\text{asym} \\
{\cal B}_{N-1} &=& \left[\tilde{y}_3 \left(\prod_{i=N-q}^{N-2} \tilde{x}_i \right) \right]^n_\text{asym} \,.
\ee
These have all the same properties as the other $N-2$ baryons.
The chiral-ring relation in this case reduces to
\begin{equation}
\label{ChiralRing4}
\prod_{i=0}^{N-1} {\cal B}_i \sim (\det {\cal M}_2)^{q} \,.
\end{equation}

\begin{figure}[ht!]
\centering
\includegraphics[scale=0.675]{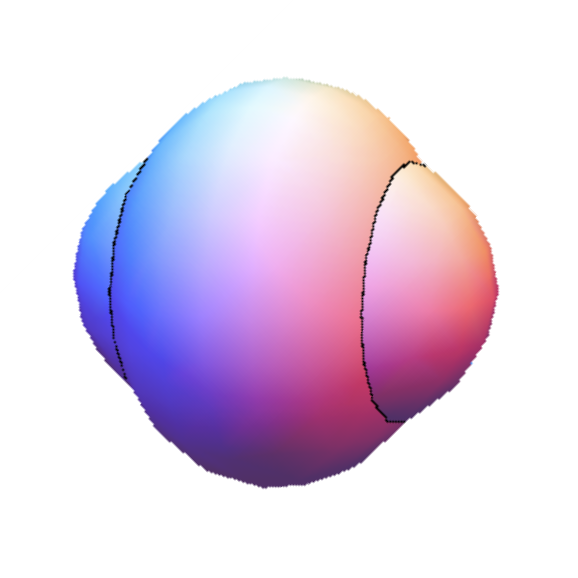} 
\caption{A two-dimensional projection of the internal space $M_3$ dual to the three-segment symmetric quiver.}
\label{SymMasFlav}
\end{figure}

The background dual to this theory has three regions, with $2\pi F_0 = -n, 0, n$, see
Fig.~\ref{SymMasFlav}.
The solution in the three regions takes the form
\begin{equation}
\alpha(z) = \begin{cases}
-\frac{(9\pi)^2}{6}z [3k(z-N)+n(3q(q-N)+z^2)] & \text{for} \; z \leq q \\[5pt]
-\frac{(9\pi)^2}{6} [nq^3-3N(k+nq)z+3(k+nq)z^2] & \text{for} \; q \leq z \leq N-q \\[5pt]
-\frac{(9\pi)^2}{6}(N-z) [-3kz+n(3q(q-N)+(N-z)^2)] & \text{for} \; z \geq N-q
\end{cases}
\end{equation}
and the RR flux on $S^2$ is given by
\be
\int_{S^2} \frac{\tilde{F}_2}{2\pi} = \left\{
\begin{array}{ll}
- k & \text{for} \; z < q \\[5pt]
- (k+nq) & \text{for} \;  q < z < N-q\\[5pt]
- k & \text{for} \; z > N-q \,. \\
\end{array}
\right.
\ee
For $k\neq 0$ there are singularities at the two poles.
As before, these are usefully described as $k$ D8-branes with $\mu_1=1$ at $z=1$ and $k$ D8-branes with $\mu_4=-1$ at $z=N-1$.
In addition there are $n$ D8-branes with $\mu_2=q$ at $z=q$, and $n$ D8-branes with $\mu_3=-q$ at $z=N-q$.
The latter are visible in Fig.~\ref{SymMasFlav}.

The non-abelian part of the symmetry is again realized directly by the D8-branes.
There are now three independent massless $U(1)$ gauge fields $A_c$ with $c=1,2,3$, parameterized by the four diagonal D8-brane $U(1)$ gauge fields
and the RR gauge field as
\be
a_b = \beta_b^{(c)} A_c\ , \quad C_1 = \gamma^{(c)} A_c \,,
\ee
with
\begin{eqnarray} 
\label{SymmetricQuiverU(1)s}
\left(
\begin{array}{l}
\beta_1^{(1,3)}\\[5pt]
\beta_2^{(1,3)}\\[5pt]
\beta_3^{(1,3)}\\[5pt]
\beta_4^{(1,3)}\\[5pt]
\gamma^{(1,3)}
\end{array} 
\right) &=& 
 \frac{1}{2(k+nq)}
 \left(
\begin{array}{c}
- n \left(q \pm \frac{k + nq}{k+n}\right)\\[5pt]
k\left(1 \pm \frac{k+nq}{k+n}\right) \\[5pt]
- k \left(1 \mp \frac{k+nq}{k+n}\right) \\[5pt]
n \left(q \mp \frac{k+nq}{k+n}\right) \\[5pt]
nk(1-q)
\end{array}
\right) \nonumber \\[5pt]
\left(
\begin{array}{l}
\beta_1^{(2)}\\[5pt]
\beta_2^{(2)}\\[5pt]
\beta_3^{(2)}\\[5pt]
\beta_4^{(2)}\\[5pt]
\gamma^{(2)}
\end{array}
\right) &=& 
\frac{n}{(k+nq)^2}
\left(
\begin{array}{c}
- \frac{nN}{2} + nq\\[5pt]
- \frac{nN}{2} - k \\[5pt]
 \frac{nN}{2} + k  \\[5pt]
 \frac{nN}{2} - nq \\[5pt]
n(k+n)\left(q-\frac{N}{2}\right)
\end{array}
\right)\,.
\end{eqnarray}
One can easily verify that these satisfy the conditions in (\ref{U(1)Condition1}) and  (\ref{U(1)Condition2}).

The three string-mesons are dual to open strings between the corresponding pair of neighboring D8-brane stacks.
The non-abelian charges under $SU(k)^2\times SU(n)^2$ are precisely those of the dual operators.
The $U(1)$ charges can be read off from (\ref{SymmetricQuiverU(1)s}), and are also 
in complete agreement with those of the dual operators (see Table~\ref{tab:BMcharges}).
For example the charges of the open string dual to ${\cal M}_1$ are
\be
Q_1(\text{F1}_{1}) &=& \frac{k}{2(k+nq)} \left(1 + \frac{k+nq}{k+n}\right) + \frac{n}{2(k+nq)}\left(p + \frac{k + nq}{k+n}\right) = 1 \\
Q_2(\text{F1}_{1}) &=&  \frac{n}{(k+nq)^2}\left(- \frac{nN}{2} - k  + \frac{nN}{2} - nq\right)  = - \frac{n}{k + nq} \\
Q_3(\text{F1}_{1})&=& \frac{k}{2(k+nq)} \left(1 - \frac{k+nq}{k+n}\right) + \frac{n}{2(k+nq)}\left(q - \frac{k + nq}{k+n}\right) = 0 \,.
\ee
The masses of the open strings are given by 
\be
M(\text{F1}_{1})&=& 2(q-1) \nonumber \\
M(\text{F1}_{2}) &=& 2(N-2q) \\
M(\text{F1}_{3}) &=& 2(q-1)\,, \nonumber
\ee
which are in agreement, for large $N$ and $q$, with the dimensions of the dual string-meson operators.

The baryon is dual to a D0-brane, possibly with $n$ strings attached if it is located in one of the two massive regions.
By symmetry it is clear that the BPS state is described by a D0-brane at the center $z=N/2$, with no strings attached.
The mass is in general given by $2r_\text{max}$, which in this case is
\be
M(\text{D0}) = 2(k+ nq) \,,
\ee
in agreement with the dimension of the baryon.
The $U(1)$ charges are given by
\be
Q_c(\text{D0}) =  \gamma^{(c)} + \frac{1}{2}\left( n\beta_3^{(c)} + k\beta_4^{(c)} - k \beta_1^{(c)} - n\beta_2^{(c)} \right) \,,
\ee
and explicitly
\be
Q_1(\text{D0})=Q_3(\text{D0}) = 0 \ , \quad Q_2(\text{D0}) = \frac{n^2}{k+nq} \,,
\ee
in complete agreement with the charges of the baryon.

As in the other examples, the chiral-ring relation (\ref{ChiralRing3}) is realized by a wrapped Euclidean D2-brane.
The NSNS flux induces $N$ units of magnetic charge that is canceled by $N$ D0-brane worldlines ending on the D2-brane.
The RR flux induces $k+qn$ units of electric charge in the massless region that is canceled by $k+qn$ worldsheets of open strings 
ending on the D2-brane in the massless region, and $k$ units of electric charge in both massive regions that are canceled
by $k$ open string worldsheets ending on the D2-brane in those regions.
The former accounts for the powers of ${\cal M}_2$ in (\ref{ChiralRing3}), and the latter for the power of ${\cal M}_1$
and ${\cal M}_3$.


\section*{Acknowledgments}

We thank F.~Apruzzi, A.~Bourget, and G.~Zafrir for useful discussions. The work of O.B.  is supported in part by the Israel Science Foundation under grant No. 1390/17. The work of M.F.~is supported in part by the Israel Science Foundation under grant No.~504/13, 1696/15, 1390/17, by the I-CORE Program of the Planning and Budgeting Committee, and by the European Union's Horizon 2020 research and innovation programme under the Marie Skłodowska-Curie grant agreement No.~754496~-~FELLINI. M.F.~wishes to thank the Weizmann Institute of Science for hospitality during the completion of this work. The work of D.R.G.~is supported in part by the Spanish government grant MINECO-16-FPA2015-63667-P, and by the Principado de Asturias through the grant FC-GRUPIN-IDI/2018/000174. A.T.~is supported in part by INFN and by the ERC Starting Grant 637844-HBQFTNCER. The authors would like to thank the GGI in Florence and the MITP in Mainz for hospitality during the completion of this work.


\appendix 

\section{Particles in global $AdS_7$}
\label{app:globalads7}

\subsection{Killing spinors in global coordinates}
\label{sub:ks}

In global coordinates, the metric of $AdS_d$ reads
\begin{equation}
ds^2_{AdS_d}=-\cosh^2 \rho d\tau^2 + d\rho^2 + \sinh^2 \rho ds^2_{S^{d-2}}\ .
\end{equation}
We can choose the Vielbein 
\begin{equation}
	e^0  =  \cosh \rho \,d \tau \, ,\quad e^a = \sinh \rho \,\tilde e^a \, ,\quad e^{d-1} = d \rho\, ,
\end{equation}
where $\tilde e^a$ ($a=1,\ldots,d-2$) is a Vielbein for the unit-radius $S^{d-2}$. In this frame, the Killing spinors read
\begin{equation}\label{eq:7dKs}
	\zeta = \exp\left[\frac12 \rho \gamma_{d-1}\right] \exp\left[\frac12 \tau \gamma_0\right] \left(\begin{array}{c}
		\zeta_+^\mathrm{S} \\ \zeta_-^\mathrm{S}
	\end{array}\right)\,,
\end{equation}           
where the gamma matrices have flat indices, and $\zeta^\mathrm{S}_\pm$ are two Killing spinors on $S^{d-2}$:
\begin{equation}
	D^\mathrm{S}_a \zeta^\mathrm{S}_\pm = \mp \frac i2 \gamma^\mathrm{S}_a\zeta^\mathrm{S}_\pm\,.
\end{equation}
An ${}^\mathrm{S}$ denotes quantities on the $S^{d-2}$.

For more explicit computations, we can choose 
\begin{equation}
	\gamma_0 = i \sigma_3 \otimes 1 \, ,\qquad \gamma_a = \sigma_1 \otimes \gamma^\mathrm{S}_a \, ,\qquad \gamma_{d-1} = \sigma_2 \otimes 1\, ,
\end{equation}
$a=1,\ldots,d-2$ being flat indices on $S^{d-2}$ and $\sigma_i$ the Pauli matrices. Let us now specialize to our case $d=7$. The Majorana conjugation matrix can be taken to be $B= \sigma_1 \otimes B^\mathrm{S}$; it obeys $B \gamma_\mu^* = \gamma_\mu B$ (and $BB^*=-1_7$), and yields the charge-conjugate spinor $\zeta^\text{c} \equiv B \zeta^*$ (with $^*$ denoting complex conjugation).

\subsection{Probe D0-branes}
\label{sub:d0prob}

We want to see if a D0 can be BPS in the $AdS_7$ solutions. If we decompose the ten-dimensional gamma matrices as 
\begin{equation}\label{eq:gamma-dec}
\Gamma_\mu= \gamma_\mu \otimes 1 \otimes \sigma_2\ , \quad \Gamma_i= 1\otimes \sigma_i \otimes \sigma_1
\end{equation}
for $\mu=0,\ldots,6$ and $i=1,2,3$, the BPS condition reads
\begin{equation}\label{eq:D0-cal}
	\Gamma_0 \epsilon_1 = \epsilon_2\,,
\end{equation} 
with $\epsilon_{1,2}$ the two supersymmetry parameters of Type IIA. These have the form \cite[(A.4)]{Apruzzi:2013yva}
\begin{equation}\label{eq:eps-ads7}
	\begin{split}
		&\epsilon_1 = (\zeta\otimes \chi_1+ \zeta^\mathrm{c} \otimes \chi_1^\mathrm{c}) \otimes v_+\,, \\
		&\epsilon_2 = (\zeta\otimes \chi_2- \zeta^\mathrm{c} \otimes \chi_2^\mathrm{c}) \otimes v_-\,, 
	\end{split}
\end{equation}
where $\sigma_3 v_\pm = \pm v_\pm$. (E.g. one can take $v_\pm = {1}/{\sqrt{2}} \left(\begin{smallmatrix} 1 \\ \mp 1 \end{smallmatrix}\right)$.) From \eqref{eq:7dKs} we compute 
\begin{subequations}
\begin{align}
	\gamma_0 \zeta &=  \exp\left[-\frac12 \rho \gamma_{d-1}\right] \exp\left[\frac12 \tau \gamma_0\right] i\sigma_3 \left(\begin{array}{c}
		\zeta_+^\mathrm{S} \\ \zeta_-^\mathrm{S}
	\end{array}\right)   
	\, ,\\
	\zeta^\mathrm{c} &=\exp\left[\frac12 \rho \gamma_{d-1}\right] \exp\left[\frac12 \tau \gamma_0\right] \sigma_1 \left(\begin{array}{c}
		\zeta_+^\mathrm{S} \\ \zeta_-^\mathrm{S}
	\end{array}\right)\,.
\end{align}
\end{subequations}
We now take $\rho=0$, since we expect a BPS particle to be static.\footnote{Every point can be mapped to any other point by an isometry; however, mapping $\rho=0$ to $\rho\neq 0$ will produce a geodesic, but not a static one.} With this, we see that we can obtain
\begin{equation}\label{eq:g0i}
	\gamma_0\zeta= i \zeta
\end{equation}
if we take $\zeta_-^\mathrm{S}=0$; its conjugate is $\gamma_0 \zeta^\mathrm{c}=-i \zeta^\mathrm{c}$. Now \eqref{eq:D0-cal} imposes 
\begin{equation}\label{eq:chi21}
	\chi_2 = -\chi_1
\end{equation}
for the spinors on the internal space $M_3$ of the ten-dimensional vacuum. From \cite[Sec.~3.2]{Apruzzi:2013yva} we find that for the $AdS_7$ solutions 
\begin{subequations}
\begin{align}
&\chi_1= e^{i(\theta_2+\theta_1)/2}\left(\cos(\psi/2) \chi+\sin(\psi/2) \chi^\mathrm{c}\right)\ , \\ 
&\chi_2= e^{i(\theta_2-\theta_1)/2}\left(\cos(\psi/2) \chi-\sin(\psi/2) \chi^\mathrm{c}\right)\ ,
\end{align}
\end{subequations}
where $\theta_1,\theta_2,\psi$ are local coordinates on the internal space. Imposing \eqref{eq:chi21} then requires
\begin{equation}\label{eq:chir2}
\theta_1 = 0 \ , \quad \psi =0\ .
\end{equation}
Chasing the redefinitions in that paper (notably \cite[(4.4), (4.8)]{Apruzzi:2013yva}), these imply $\beta=0$, which denotes the north pole of the $S^2$ fiber, and $x=0$. In the language of the present paper,
\begin{equation}
\beta \equiv \theta\ , \quad x \equiv \frac{\dot{\alpha}}{(\dot \alpha^2 - 2 \alpha \ddot \alpha)^{1/2}} \ ,
\end{equation}
where the function $x$ approaches $-1$ at $z=0$ and $1$ at $z=N$. Then \eqref{eq:chir2} says $\theta =0$ and $\dot \alpha=0$.

All in all, we have found that a D0 can be BPS at the locus
\begin{equation}
	\rho=0 \, ,\quad \dot \alpha=0 \, ,\quad \theta=0\,,
\end{equation}
where the latter denotes the north pole of the $S^2$.

\subsection{Probe F1-strings}
\label{sub:f1prob}

We now consider fundamental strings stretched along the $z$ direction. The BPS condition reads
\begin{equation}
	\Gamma_{0\hat z} \epsilon_1 = \epsilon_1 \ , \quad \Gamma_{0\hat z} \epsilon_2 =- \epsilon_2\ ,
\end{equation}
where $\hat z$ denotes the flat index corresponding to the $z$ direction. Using again \eqref{eq:gamma-dec}, \eqref{eq:eps-ads7}, (\ref{eq:g0i}), this becomes
\begin{equation}\label{eq:gammaz-chi}
	\gamma_{\hat z} \chi_a = \chi_a
\end{equation}
for both $\chi_a$. 

It is now convenient to use the expression for the spinors given in \cite[Sec.~3.1]{Rota:2015aoa}:\footnote{It is also possible to use these to analyze (\ref{eq:chi21}) in the previous subsection, of course.} 
\begin{equation}
	\chi_1 = \exp\left[-\frac i2\left(\psi + \frac\pi2\right)\right] \chi_{S^2}
	\, ,\quad
	\chi_2 = i \exp\left[\frac i2\left(\psi - \frac\pi2\right)\right] \chi_{S^2}\,,
\end{equation}
where $\chi_{S^2}$ is the Killing spinor on the internal $S^2$. This is written in a frame where $\gamma_{\hat z}=\sigma_3$, which is also the chiral gamma in the $S^2$ direction. With this, we see that (\ref{eq:gammaz-chi}) requires the component of $\chi_{S^2}$ with negative chirality to vanish. This happens at the north pole of the $S^2$. 

Thus we have found that an F1 stretched along $z$ can be BPS at the locus $\theta=0$.

\section{Abelian charges of operators and dual string states}
\label{app:charges}

In this appendix we identify a consistent set of numbers $\{\beta_b^{(c)},\gamma^{(c)}\}$ such that the $U(1)$ charges of the open string given in \eqref{eq:F1bb} coincide with those of the string-meson given in \eqref{StringMesonCharges}, and the $U(1)$ charge of the D0-brane-strings combination given in \eqref{eq:D0F1U1charge} coincides with that of the baryon given in \eqref{BaryonChargesPlateau}.

For later convenience, let us repeat here our conventions. In the generic brane configuration with plateau, there are $s\geq 0$ D8 stacks (i.e. flavor nodes) labeled by $b=1,\ldots,s$, each containing $n_b >0$ branes and located at $z=j=j_b \in [0,N]$ along the base interval. (Remember that the $n_b$ may not be all independent, since once we specify all the ranks $r_j$ then we must have $n_b = 2r_{j_b} - r_{j_b-1} - r_{j_b+1}$.) Accordingly we can define $s-1$ string-mesons $\mathcal{M}_b$ in field theory, beginning in the $b$th and ending in the $(b+1)$st flavor. 

The plateau is the subinterval $[j_p,j_{p+1}] \subset [0,N]$, i.e. it begins at the last D8 stack in the left region located at $j=j_p$, and ends at the first stack on the right region located at $j=j_{p+1}$.

\subsection{Field theory}

The $s-1$ $U(1)$ charges of the $s-1$ string-mesons are:
\begin{equation}
Q_c(\mathcal{M}_b) =\begin{cases} \displaystyle
 - \frac{n_{b-1}}{r_{j_b}}(1-\delta_{b,1})  & \text{for}\ c=b-1 \\
\displaystyle  \frac{n_{b+1}}{r_{j_b}} + \sum_{k=j_b}^{j_{b+1}-1} \frac{n_b n_{b+1}}{r_k r_{k+1}} + \frac{n_b}{r_{j_{b+1}}}  &\text{for}\ c=b \\
 \displaystyle - \frac{n_{b+2}}{r_{j_{b+1}}}(1-\delta_{b,s-1}) \displaystyle & \text{for}\ c=b+1\\
 0 & \text{otherwise}
 \end{cases}
\end{equation}
The Kronecker $\delta$'s are there to enforce the absence of a contribution for $b=1$ and $b=s-1$, as appropriate. Compactly:
\begin{multline}\label{eq:mesons}
Q_c({\cal M}_b) = -(1-\delta_{b,1}) \delta_{c,b-1} \frac{n_{b-1}}{r_{j_b}} -(1-\delta_{b,s-1}) \delta_{c,b+1}\frac{n_{b+2}}{r_{j_{b+1}}}\ + \\ + \delta_{c,b}\left( \frac{n_{b+1}}{r_{j_b}} + \sum_{k=j_b}^{j_{b+1}-1} \frac{n_b n_{b+1}}{r_k r_{k+1}} + \frac{n_b}{r_{j_{b+1}}}  \right)\ .
\end{multline}

On the other hand there are $N$ baryons, which are all identified in the SCFT. Its $s-1$ $U(1)$ charges are
\begin{equation}\label{eq:baryon}
Q_c(\mathcal{B}) = \delta_{c,p}\frac{n_{p} n_{p+1}}{r_\text{max}}\ ,
\end{equation}
namely the baryon is charged only under the $p$th $U(1)$.

\subsection{Supergravity}

In supergravity the $s-1$ $U(1)$ charges of the F1-string stretched between the $b$th and $(b+1)$st D8 stack are given by
\begin{equation}\label{eq:F1}
Q_c(\text{F1}_b) = \beta_{b+1}^{(c)} - \beta_{b}^{(c)} \ .
\end{equation}
The $U(1)$ charges of the D0-brane-strings combination inserted in the plateau $j_p < j < j_{p+1}$ are given by
\begin{equation}\label{eq:D0}
Q_c(\text{D0}+\text{F1}) = \gamma^{(c)}  -  \sum_{b=1}^{p} n_b \beta_b^{(c)} \ .
\end{equation}
From the supergravity analysis we know the massless combinations of $U(1)$'s have to satisfy the constraints
\begin{equation}\label{eq:constraints}
\sum_{b=1}^s n_b \beta_b^{(c)} = 0 \ , \quad N \gamma^{(c)} + \sum_{b=1}^s j_b n_b \beta_b^{(c)} = 0\ .
\end{equation}


\subsection{Solving the constraints}

The strategy is to solve for the $\beta_b^{(c)}$. We have $s$ of them but one is linearly determined by the first constraint in \eqref{eq:constraints} in terms of the others, so we are left with $s-1$. Then we need to solve the $(s-1)^2$ equations
\begin{align}
\beta_{b+1}^{(c)} - \beta_{b}^{(c)} &= -(1-\delta_{b,1}) \delta_{c,b-1} \frac{n_{b-1}}{r_{j_b}} -(1-\delta_{b,s-1}) \delta_{c,b+1}\frac{n_{b+2}}{r_{j_{b+1}}} \ + \nonumber \\
& \ \ \ \, +\delta_{c,b}\left( \frac{n_{b+1}}{r_{j_b}} + \sum_{k=j_b}^{j_{b+1}-1} \frac{n_b n_{b+1}}{r_k r_{k+1}} + \frac{n_b}{r_{j_{b+1}}}  \right) \equiv f_{b,c}\ .
\end{align}
Each of these is an inhomogeneous first-order difference equation. Once solved together, these provide the $\beta_b^{(c)}$ in terms of an ``integration constant'', say $\beta_1^{(c)}$, which we determine via the first constraint in \eqref{eq:constraints}. 

This then yields $\gamma^{(c)}$ given the second constraint in \eqref{eq:constraints}. Finally we need to check that, by plugging the expression for $\gamma^{(c)}$ and $\beta_{b}^{(c)}$ for $b=1,\ldots,p$ we just determined into \eqref{eq:D0}, we get the number on the RHS of \eqref{eq:baryon} (which constitutes a nontrivial test of the duality). This then represents a consistent set of charges $\{\beta_b^{(c)},\gamma^{(c)}\}$ for the probe F1's and D0 inserted in the gravity dual.


Let us sum the above equations from $1$ to $b-1$:
\begin{equation}
\sum_{k=1}^{b-1} \beta_{k+1}^{(c)}-\beta_{k}^{(c)} = \beta_b^{(c)} - \beta_1^{(c)} = \sum_{k=1}^{b-1} f_{k,c}\ , \quad b=2,\ldots,s\ .
\end{equation}
Therefore
\begin{equation}
\beta_b^{(c)} = \beta_1^{(c)} + \sum_{k=1}^{b-1} f_{k,c} \ , \quad b=2,\ldots,s\ .
\end{equation}
On top of this we must impose
\begin{equation}\label{eq:cons1}
0 = \sum_{b=1}^{s} n_b \beta_b^{(c)} = n_1 \beta_1^{(c)} +  \sum_{b=2}^{s} n_b \beta_b^{(c)}=  n_1 \beta_1^{(c)}  + \sum_{b=2}^{s} n_b \left(\beta_1^{(c)} + \sum_{k=1}^{b-1} f_{k,c}  \right) \ ,
\end{equation}
so that 
\begin{equation}\label{eq:beta1}
\beta_1^{(c)} \ = - \frac{\sum_{m=2}^{s} n_m ( \sum_{k=1}^{m-1}f_{k,c})}{\sum_{k=1}^{s} n_k}\equiv - \tau_c \ .
\end{equation}
Therefore
\begin{equation}
\beta_b^{(c)} = \sum_{k=1}^{b-1} f_{k,c}  - \frac{\sum_{m=2}^{s} n_m ( \sum_{k=1}^{m-1}f_{k,c})}{\sum_{k=1}^{s} n_k} \equiv \sigma_{b,c} -  \tau_c \ , \quad b=2,\ldots,s\ ,
\end{equation}
having defined the sum $\sigma_{b,c} \equiv \sum_{k=1}^{b-1} f_{k,c}$ for $b=2,\ldots,s$. Notice that only the first summand  in the above line depends on $b$, whereas the second one, $\tau_c$ for $c=1,\ldots,s-1$, is common to all $\beta^{(c)}$'s. To have explicit expressions we need to evaluate $\sigma_{b,c}$ for $b=2,\ldots,s$, as well as the nested sum $\sum_{m=2}^{s} n_m \sigma_{m,c}$. Finally, given \eqref{eq:beta1}, we can extend the above expression to $b=1$ by writing
\begin{equation}\label{eq:beta-final}
\beta_b^{(c)} =  (1-\delta_{b,1})\sigma_{b,c} - \tau_c\  , \quad b=1,\ldots,s\ , \ c=1,\ldots,s-1\ ,
\end{equation}
which provides a closed expression valid for all $\beta$'s.

Now we need to check that $Q_c(\text{D0}+\text{F1})$ in \eqref{eq:D0} gives $Q_c(\mathcal{B}) = \delta_{c,p} \frac{n_p n_{p+1}}{r_\text{max}}$; namely that
\begin{align}
Q_c(\text{D0}+\text{F1}) &= \gamma^{(c)}  -  \sum_{b=1}^{p} n_b \beta_b^{(c)} = \tau_c \sum_{b=p+1}^s n_b -\frac{1}{N} \sum_{b=2}^s j_b n_b \sigma_{b,c} -  \sum_{b=1}^{p} n_b \beta_b^{(c)} \nonumber \\
&= \tau_c \sum_{b=1}^s n_b - \sum_{b=2}^p n_b \sigma_{b,c}  -\frac{1}{N} \sum_{b=2}^s j_b n_b \sigma_{b,c} \nonumber \\ 
&=  \sum_{b=2}^s n_b \sigma_{b,c} - \sum_{b=2}^p n_b \sigma_{b,c}  -\frac{1}{N} \sum_{b=2}^s j_b n_b \sigma_{b,c} \label{eq:LHS}\\
&\overset{!}{=} \delta_{c,p} \frac{n_p n_{p+1}}{r_\text{max}} = \delta_{c,p} \frac{n_p n_{p+1}}{\sum_{b=1}^p j_b n_b} = \delta_{c,p} \frac{n_p n_{p+1}}{\sum_{b=p+1}^s (N-j_b) n_b} = Q_c(\mathcal{B})\ . \label{eq:RHS}
\end{align}
To go from the first to the second line we used the definition of $\beta_b^{(c)}$ for $b=1,\ldots,s$ given in \eqref{eq:beta-final}; to go from the second to the third we used the definition of $\tau_c$ given in \eqref{eq:beta1}. If $s=2$ then necessarily $p=1$ ($p<s$ by construction), and we simply drop the second sum in the penultimate line. Even if $s>2$, if $p=1$ we drop that sum. Hence, for $c=1,\ldots,s-1$:
\begin{align}
Q_c(\text{D0}+\text{F1}) &= \begin{cases} \displaystyle \sum_{b=2}^s n_b \sigma_{b,c}-\frac{1}{N} \sum_{b=2}^s j_b n_b \sigma_{b,c} & \text{for}\ p=1 \\
 \displaystyle \sum_{b=p+1}^s n_b \sigma_{b,c}-\frac{1}{N} \sum_{b=2}^s j_b n_b \sigma_{b,c} & \text{for}\ 1 \neq p <s \end{cases} \nonumber \\
 &\overset{!}{=} \delta_{c,p} \frac{n_p n_{p+1}}{r_\text{max}}  = Q_c(\mathcal{B})\ .
\end{align}
In the following subsections we will prove the above identity in the $s=2$ and $s=3$ cases for illustrative purposes.

\subsubsection{The simplest case of $s=2$}

Let us compute $\sigma_{b,c}$ and $\tau_c$, with $b=2,\ldots,s$ and $c=1,\ldots,s-1$ in the simplest case, namely $s=2$ (hence $p=1$). The plateau is $[j_1,j_2] \subset [0,N]$, which is $(j_2-j_1)$-long, and the ranks are all constant there: $r_k = r_\text{max}$ for $k=j_1,\ldots,j_2$. In this case there are only two sums, $\sigma_{2,1}=f_{1,1}$ and $\tau_1 = \frac{n_2\, \sigma_{2,1}}{n_1+n_2}=\frac{n_2\, f_{1,1}}{n_1+n_2}$, so we only need to compute $f_{1,1}$:
\begin{subequations}\label{eq:sigma}
\begin{align}
\sigma_{2,1} &= f_{1,1} = \frac{n_{2}}{r_{j_1}} + \sum_{k=j_1}^{j_{2}-1} \frac{n_1 n_{2}}{r_k r_{k+1}} + \frac{n_1}{r_{j_{2}}} =\frac{n_{2}}{r_\text{max}} + \frac{n_1}{r_\text{max}} + n_1 n_2 (j_2-j_1) \frac{1}{r_\text{max}^2} \nonumber \\
& = \frac{r_\text{max}(n_1+n_2) + (j_2-j_1)n_1 n_2}{r_\text{max}^2}\ ; \\
\tau_1 &=  \frac{n_2}{n_1+n_2} f_{1,1} = \frac{n_2}{n_1+n_2} \frac{r_\text{max}(n_1+n_2) + (j_2-j_1)n_1 n_2}{r_\text{max}^2}\ .
\end{align}
\end{subequations}
With this, let us evaluate \eqref{eq:LHS}:
\begin{align}
&\tau_1 n_2 -\frac{1}{N}j_2 n_2 \sigma_{2,1} - n_1 \beta_1^{(1)}  = n_2 f_{1,1} -\frac{1}{N} (N n_2-r_\text{max}) f_{1,1} =   \frac{r_\text{max}}{N} f_{1,1}\ .
\end{align}
Remembering \eqref{eq:sigma}, we obtain
\begin{align}
& \frac{r_\text{max}(n_1+n_2) + (j_2-j_1)n_1 n_2}{N r_\text{max}} =\frac{n_2}{r_\text{max}+j_2 n_2} \frac{r_\text{max}(n_1+n_2) + (j_2-j_1)n_1 n_2}{r_\text{max}}\ .
\end{align}
Upon plugging in $r_\text{max}=n_1 j_1$, the above expression reduces to $\frac{n_1 n_2}{r_\text{max}}$, i.e. \eqref{eq:RHS}, as expected.

\subsubsection{The case of $s=3$}

For $s=3$ we have $b=2,3$ and $c=1,2$. The endpoints of the plateau are now located at $j_2 = j_p$ and $j_3 = j_{p+1}$ (i.e. we picked $p=2\neq 1$). We then place the first stack on the left of the plateau, at $j_1$, and the nearest stack on its right is at $j_2$, i.e. $a_\text{R}(j_1)=j_2$. 

We need to compute $\sigma_{b,c}$ and $\tau_c$:
\begin{align}
&\sigma_{2,c} = f_{1,c} \ , \quad \sigma_{3,c} = f_{1,c} + f_{2,c} \ ; \quad \tau_c = \frac{n_2 \sigma_{2,c}+n_3 \sigma_{3,c}}{n_1+n_2+n_3} =  \frac{(n_2+n_3) f_{1,c} + n_3 f_{2,c}}{n_1+n_2+n_3}\ .
\end{align}
In turn the $f_{b,c}$ are given by:
\begin{subequations}
\begin{align}
f_{1,1} & = \frac{n_2}{r_{j_1}} + \sum_{k=j_1}^{j_2 -1} \frac{n_1 n_2}{r_k r_{k+1}} + \frac{n_1}{r_{j_2}} = \frac{n_2}{r_{j_1}} + \frac{n_1}{r_\text{max}} + \sum_{k=j_1}^{j_2 -1} \frac{n_1 n_2}{r_k r_{k+1}}\nonumber \\
&=  \frac{{n_1}+{n_2}}{({j_1} -{j_2}) {n_2}+r_\text{max}} = \frac{1}{j_1}\ ,\\
f_{1,2} & = - \frac{n_3}{r_{j_2}}  = - \frac{n_3}{r_\text{max}} \ , \\
f_{2,1} & = - \frac{n_1}{r_{j_2}} =   - \frac{n_1}{r_\text{max}}\ , \\
f_{2,2} & = \frac{n_3}{r_{j_2}} + \sum_{k=j_2}^{j_3 -1} \frac{n_2 n_3}{r_k r_{k+1}} + \frac{n_2}{r_{j_3}} = \frac{n_{3}}{r_\text{max}} + \frac{n_2}{r_\text{max}} + n_2 n_3 (j_3-j_2) \frac{1}{r_\text{max}^2} \nonumber \\
& = \frac{r_\text{max}(n_2+n_3) + (j_3-j_2)n_2 n_3}{r_\text{max}^2} = \frac{{j_1} {n_1} {n_3}+{j_3} {n_2} {n_3}+{n_2} r_\text{max}}{r_\text{max}^2}\ .
\end{align}
\end{subequations}
We used the identities $r_\text{max} = j_1n_1+j_2 n_2 = (N-j_3)n_3$ and the fact $r_j =r_\text{max}-(j_2-j)n_2$ for all $j<j_2$ and $r_j = r_\text{max}$ for all $j_2 \leq j \leq j_3$. Therefore
\begin{subequations}
\begin{align}
Q_c(\text{D0}+\text{F1}) &= \sum_{b=3}^3 n_b \sigma_{b,c}-\frac{1}{N} \sum_{b=2}^s j_b n_b \sigma_{b,c}  \\
& = n_3 (f_{1,c}+f_{2,c}) -\frac{j_2 n_2 +Nn_3 -r_\text{max}}{N} f_{1,c} -\frac{N n_3 -r_\text{max}}{N} f_{2,c} \nonumber \\
&=  \frac{r_\text{max}-j_2 n_2}{N} f_{1,c} +\frac{r_\text{max}}{N} f_{2,c} = \frac{n_3 r_\text{max} (f_{1,c}+f_{2,c}) - j_2 n_2 n_3 f_{1,c}}{r_\text{max}+j_3n_3}\nonumber \\
&=   \frac{n_3 r_\text{max} (f_{1,c}+f_{2,c}) - j_2 n_2 n_3 f_{1,c}}{r_\text{max}+j_3n_3} \\
&\overset{!}{=} Q_c (\mathcal{B}) =  \delta_{c,2} \frac{n_2 n_3}{r_\text{max}}\ .
\end{align}
\end{subequations}
We need to check that the above identity is satisfied for both $c=1$ and $c=2$; namely that
\begin{equation}
 \frac{n_3 r_\text{max} \left( \frac{1}{j_1} - \frac{n_1}{r_\text{max}} \right) - \frac{j_2 n_2 n_3}{j_1}}{r_\text{max}+j_3n_3} = 0\ ,
\end{equation}
and
\begin{equation}
 \frac{n_3 r_\text{max} \left(-\frac{n_3}{r_\text{max}} +  \frac{r_\text{max}(n_2+n_3) + (j_3-j_2)n_2 n_3}{r_\text{max}^2}  \right) + j_2 n_2 n_3 \frac{n_3}{r_\text{max}}}{r_\text{max}+j_3n_3} = \frac{n_2 n_3}{r_\text{max}}\ .
\end{equation}
The first is satisfied upon using $r_\text{max} = j_1n_1+j_2 n_2$, the second is automatically satisfied without using any further identity.


\end{document}